\documentclass[aps,showpacs,preprint,superscriptaddress]{revtex4}
\usepackage{graphicx}
\usepackage{subfigure}
\usepackage{float}
\usepackage{amsmath}
\usepackage{amsfonts}
\usepackage{bm}
\usepackage{bbm}
\usepackage{txfonts}
\usepackage{array}
\usepackage{amssymb}

\begin{document}

\title{Effects of polarization and high harmonics of two-color fields on dynamically assisted pair production}
\author{Obulkasim Olugh}
\affiliation{Key Laboratory of Beam Technology of the Ministry of Education,
and College of Nuclear Science and Technology, Beijing Normal University, Beijing 100875, China}
\author{Zi-Liang Li}
\affiliation{School of Science, China University of Mining and Technology, Beijing 100083, China}
\author{Bai-Song Xie \footnote{bsxie@bnu.edu.cn}}
\affiliation{Key Laboratory of Beam Technology of the Ministry of Education,
and College of Nuclear Science and Technology, Beijing Normal University, Beijing 100875, China}
\affiliation{Beijing Radiation Center, Beijing 100875, China}

\date{\today}
\begin{abstract}
Electron-positron pair production in dynamically assisted two-color electric fields is investigated for various polarizations. Momentum spectrum and number density of the created pairs are examined carefully, in particular, the effects of polarization and high harmonics of two-color fields are exhibited. For only single strong field, the interference effects of momentum spectrum would vanish when polarization is high, however, for the dynamical assisted two-color fields, the interference effects would be more and more remarkable with polarization. The multiple peaks of momentum spectrum in elliptic or/and circular polarization are observed and explained. It is found that there exists a strong nonlinear dependence of the number density on the high harmonics of two-color fields, for example, the number density can be enhanced significantly over $7-8$ orders when appropriate high harmonics is present. Another interesting finding is that the polarization effect on number density is gradually weaken as the high harmonics increases, however, a weak nonlinearity relation appears again if the high harmonics exceeds a single-photon threshold.
\end{abstract}
\pacs{12.20.Ds, 03.65.Pm, 02.60.-x}
\maketitle

\section{Introduction}

A strong background electric field causes the quantum electrodynamic (QED) vacuum a decay
accompanied by the electron-positron ($e^{-}e^{+}$) pair production, which is known as the Sauter-Schwinger effect \cite{Sauter:1931zz,Heisenberg:1935qt,Schwinger:1951nm}. This remarkable prediction of QED is a nonperturbative process that has not been experimentally observed yet \cite{Gelis:2015kya} because the pair production rate $\exp (-\pi E_{cr} /E )$ is exponentially suppressed for the electric field $E$ smaller than the Schwinger critical field strength $E_{cr} =  {m_e^2c^3} / {e\hbar} = 1.3 \times 10^{16}\rm{V/cm}$, where the corresponding critical laser intensity $I_{cr}=4.3 \times10^{29} \rm{W/cm^2}$. On the other hand, however, the high intense electromagnetic field can be achieved by high power lasers, and with recent advance of high-intensity laser technology \cite{Heinzl:2008an,Marklund:2008gj,Pike:2014wha}, the laser intensity of the order of $10^{26}$W/cm$^{2}$ are expected by Extreme Light Infrastructure (ELI) \cite{ELI} in current construction; and in the planned facilities as the Exawatt Center for Extreme Light Studies (XCELS), the Station of Extreme Light at the Shanghai Coherent Light Source, one may expect the experimental tests of Sauter-Schwinger effects will be feasible in the near future. Meanwhile, the already operating X-ray free electron laser (XFEL) at DESY in Hamburg can get
near-critical field strength as large as $E\approx 0.1\, E_{cr}$ \cite{Ringwald:2001ib}, which is enough to produce a considerable number of $e^{-}e^{+}$ pairs \cite{Alkofer:2001ik,Roberts:2002py}.

Beside a nonperturbative characteristic, Sauter-Schwinger pair production is also a nonequilibrium process with a typical non-Markovian effect \cite{Schmidt}. Over past a few decades, many theoretical works have been performed based on a number of different theoretical methods to cope with such difficulties, more details can be seen in review \cite{Gelis:2015kya,Xie}. Among them a dynamically assisted Schwinger mechanism, which combines two laser fields with low-frequency strong field and high-frequency weak field, was proposed by Sch\"utzhold \emph{et al.} \cite{Schutzhold:2008pz}, where the pair production rate is increased significantly due to lowering the Schwinger critical limit by $2$-$3$ orders of magnitude. From then on, many interesting findings in pair production researches have been achieved \cite{Orthaber:2011cm,King,Nousch:2012xe,Akal,Schneider:2016vrl,Torgrimsson:2016ant,Sitiwaldi:2018wad,Aleksandrov}. Two important features of studies on the $e^{-}e^{+}$ production are impressive recently. One is that with more realistic field parameters by choosing an appropriate pulse shape. The other is that the polarized effect has to be considered for the studied problem. Experimentally, for example, a polarization of up to $\pm0.93$ is already achieved \cite{Pfeiffer1} for low-intensity electric fields. Moreover, while the number density of created pairs is the main concern on the study, the momentum spectra of pairs  could be helpful for understanding the dynamics of the problem in point of view theoretically as well as experimentally.

Therefore, motivated by these advances and interests, in this work, we shall extend involved study by considering a realistic laser electric field with envelop pulse and without perfect linear/circular polarization. The main research is focused on the polarized electric field effects on dynamically assisted pair production. The real-time Dirac-Heisenberg-Wigner (DHW) formalism \cite{Vasak:1987um,Hebenstreit:2010vz,Hebenstreit:2011pm} is adapted, as a very efficient theoretical approach which has been used extensively for numerical calculations of pair production, for instances, in rotating circularly polarized electric fields \cite{Blinne:2013via,Blinne:2016yzv,Fillion:2017,Kohlfurst:2018kxg}, elliptic polarized background  fields \cite{Li:2015cea} and also recent work for the frequency chirp effects in differently polarized fields \cite{olugh}. In addition, it is noted that the field ellipticity effects on pair production has also been studied by other approaches beside DHW \cite{Xie:2012,Krajewska:2012,Wollert:2015} for plane-wave fields as well as for time-dependent electric fields.

We focus on the study of $e^{-}e^{+}$ pair production in dynamically assisted fields with various polarizations.
The electric field is considered as the combination of strong but slowly varying field, $\mathbf{E}_{1s}(t)$ and a weak but rapidly changing
field, $\mathbf{E}_{2w}(t)$. So the explicit form of the external field, $\mathbf{E}(t)=\mathbf{E}_{1s}(t)+\mathbf{E}_{2w}(t)$, is given as
\begin{equation}\label{eq1}
\mathbf{E}(t) \,\, =\, \, \frac{E_{1s}}{\sqrt{1+\delta^{2}}}\,
\exp\left(-\frac{t^2}{2\tau^2}\right) \,
\left(\begin{array}{c}
          \cos(\omega t+\varphi) \\
           \delta\sin(\omega t+\varphi) \\
              0 \\
        \end{array}\right)+
        \, \, \frac{E_{2w}}{\sqrt{1+\delta^{2}}}\,
\exp\left(-\frac{t^2}{2\tau^2}\right) \,
\left(\begin{array}{c}
          \cos(b \omega t+\varphi) \\
           \delta\sin(b \omega t+\varphi) \\
              0 \\
        \end{array}\right),
\end{equation}
where $E_{1s,2w}/\sqrt{1+\delta^2}$ is the amplitude of the electric field, $\tau$ denotes the envelop pulse length, $\omega$ denotes the oscillating frequencies, $\varphi$ is the carrier phase, $b$ is the high harmonics order of the weak field to strong field and $\mid\delta\mid\leq1$ represents the field polarization (or the ellipticity). In order to provide a clean multiphoton signal we chooses $b\geq9$. Such purely time dependent electric field Eq.\eqref{eq1} could be considered as the approximation of standing wave formed by two coherent contour propagating laser beams with the effects of spatial part neglected for each different polarization.
In this study the characteristic fixed field parameters are chosen as: $E_{1s} = 0.1 \sqrt{2}E_{cr}$, $E_{2w} = 0.01 \sqrt{2}E_{cr}$, $\omega=0.05m$ and $\tau=100/m$, where $m$ is the electron mass. Throughout this paper natural units of $\hbar=c=1$ are used.
When the weak but rapidly changing field $\mathbf{E}_{2w}(t)$ is set to zero,
the electric field Eq.\eqref{eq1} reduces to the one studied in \cite{Li:2015cea}.

\begin{figure}[hb]
\begin{center}
\includegraphics[width=15cm, height=10cm]{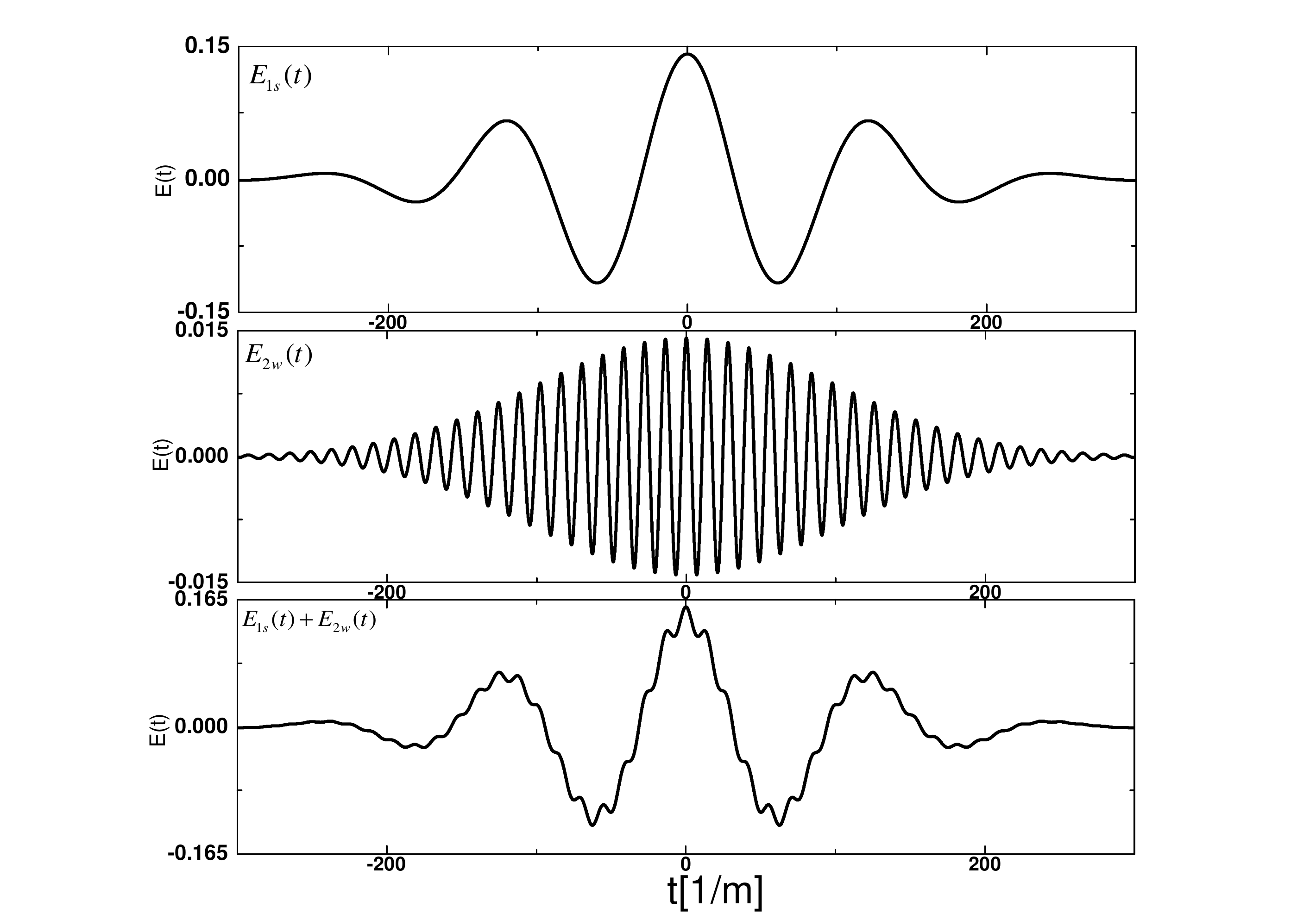}
\end{center}
\caption{The time dependence of the electric field $E(t)$ in units of the critical
field for the linearly polarized ($\delta=0$) case.
The chosen parameters are $E_{1s}=0.1\sqrt{2}E_{cr}$, $E_{2w}=0.01\sqrt{2}E_{cr}$, $\omega=0.05m$, and $\tau=100/m$
where $m$ is the electron mass.
The upper panel is for the single strong field $E_{1s}(t)$ and the middle is for the weak field $E_{2w}(t)$ when $b=9$. The lower panel displays for the dynamically assisted field $E(t)$ with a high harmonics $b\omega=0.45m$.}
\label{fig:E(t)}
\end{figure}

It is well known that the Keldysh adiabaticity parameter is defined as $\gamma=m\omega/eE$ \cite{Keldysh}, where $E$ and $\omega$
are the strength and frequency of the external electric field, such that the Schwinger (tunneling) and multiphoton pair creation can be characterised by $\gamma\ll1$ and $\gamma\gg1$, respectively. For the given parameters in this paper the Keldysh adiabaticity parameters are thus $\gamma_{1s}= 0.35 \sqrt{1+\delta^{2}}$ and $\gamma_{2w}= 3.55b \sqrt{1+\delta^{2}}$ for a single either strong or weak field, respectively.

A typical case of how the high harmonics $b\omega$ affects the time depended electric fields is displayed in Fig.~\ref{fig:E(t)}(lower panel) for linear polarization $\delta=0$ when $b=9$. For the high harmonics $b\omega$, we examined several cases in the interval $0.45m\le b\omega \le 2.5m$, and for the polarization we choose four different $\delta$ as typical studied situations. By the way, we are aware that in some circumstances the high harmonics $b\omega$ is too large to be lies in the normal regime. However, in this study, since our main interest is the influence of the high harmonics on the momentum spectrum and pair production rate for different polarization, it is valuable to examine the effect of a large $b\omega$ on number density, which corresponds to the single photon absorption or even above the threshold $2m$.

In the following, by using DHW formalism, we numerically compute the momentum spectrum and the number density of the produced pair for several values of the high harmonics $b\omega$ of laser pulses when the carrier phase is chosen as $\varphi=0$ for a set of typical polarized situation.

This paper is organized as follows. In Sec.\ref{method}, we introduce briefly the DHW formalism as well as WKB approximation which is used in our calculation for completeness. In Sec.\ref{result1}, we show the numerical results for momentum spectra and analyze the underlying physics. In Sec.~\ref{result2}, we present our numerical results for the number densities for different high harmonics and different polarizations. We end our paper with a brief conclusion in the last section.

\section{A brief outline on DHW formalism and WKB approximation}\label{method}

The DHW formalism is a relativistic phase-space quantum kinetic approach \cite{Vasak:1987um} that has been now widely adopted to study the pair production from QED vacuum in strong background field.
In the following, we present a brief review of the DHW formalism.

A convenient starting point is the gauge-invariant density operator of two Dirac field operators in the Heisenberg picture
\begin{equation}\label{density}
\hat {\mathcal C}_{\alpha \beta} \left( r , s \right) = \mathcal U \left(A,r,s
\right) \ \left[ \bar \psi_\beta \left( r - s/2 \right), \psi_\alpha \left( r +
s/2 \right) \right],
\end{equation}
in terms of the electron's spinor-valued Dirac field $\psi_\alpha (x)$, where
$r$ denotes the center-of-mass and $s$ the relative coordinates, respectively. The Wilson-line factor before the commutators
\begin{equation}
 \mathcal U \left(A,r,s \right) = \exp \left( \mathrm{i} \ e \ s \int_{-1/2}^{1/2} d
\xi \ A \left(r+ \xi s \right)  \right)
\end{equation}
 is used to keep the density operator gauge-invariant, and this factor depends on the
elementary charge $e$ and the background gauge field $A$, respectively.  In addition, we use a mean-field (Hartree) approximation via replacing gage field operator by background field.

 The important quantity of the DHW  method is the covariant Wigner operator given as the Fourier transform of the density operator \eqref{density},
\begin{equation}
 \hat{\mathcal W}_{\alpha \beta} \left( r , p \right) = \frac{1}{2} \int d^4 s \
\mathrm{e}^{\mathrm{i} ps} \  \hat{\mathcal C}_{\alpha \beta} \left( r , s
\right),
\end{equation}
and taking the vacuum expectation value of the Wigner operator gives the Wigner function
\begin{equation}
 \mathbbm{W} \left( r,p \right) = \langle \Phi \vert \hat{\mathcal W} \left( r,p
\right) \vert \Phi \rangle.
\end{equation}
By decomposing the Wigner function in terms of a complete basis set 
of Dirac matrices, we can get 16 covariant real Wigner components
\begin{equation}
\mathbbm{W} = \frac{1}{4} \left( \mathbbm{1} \mathbbm{S} + \textrm{i} \gamma_5
\mathbbm{P} + \gamma^{\mu} \mathbbm{V}_{\mu} + \gamma^{\mu} \gamma_5
\mathbbm{A}_{\mu} + \sigma^{\mu \nu} \mathbbm{T}_{\mu \nu} \right) \, .
\label{decomp}
\end{equation}
According to the  Ref. \cite{Hebenstreit:2010vz,Hebenstreit:2011pm} the equations of motion for the Wigner function are
\begin{equation}
D_{t}\mathbbm{W} = -\frac{1}{2}\vec{D}_{\vec{x}}[\gamma^{0}\vec{\gamma},\mathbbm{W}]
+im[\gamma^{0},\mathbbm{W}]-i\vec{P}\{\gamma^{0}\vec{\gamma},\mathbbm{W}\},
\label{motion}
\end{equation}
where $D_{t}$, $\vec{D}_{\vec{x}}$ and $\vec{P}$ denote the pseudodifferential operators
\begin{equation}
\begin{array}{l}
D_{t}=\partial_{t}+e\int^{1/2}_{-1/2}d\lambda \vec{E}(\vec{x}+i\lambda\vec{\bigtriangledown}_{\vec{p}},t)\cdot\vec{\bigtriangledown}_{\vec{p}},\\
\vec{D}_{\vec{x}}=\vec{\bigtriangledown}_{\vec{x}}+e\int^{1/2}_{-1/2}d\lambda \vec{B}(\vec{x}+i\lambda\vec{\bigtriangledown}_{\vec{p}},t)\times\vec{\bigtriangledown}_{\vec{p}},\\
\vec{P}=\vec{p}-ie\int^{1/2}_{-1/2}d\lambda \lambda\vec{B}(\vec{x}+i\lambda\vec{\bigtriangledown}_{\vec{p}},t)\times\vec{\bigtriangledown}_{\vec{p}}.
\end{array}\label{eq2}
\end{equation}

 Inserting the decomposition Eq. \eqref{decomp} into the equation of motion Eq. \eqref{motion} for the Wigner function, one can obtain a system of partial differential equations(PDEs) for the 16 Wigner components.
  Furthermore, for the spatially homogeneous electric fields like Eq. \eqref{eq1}, by using the characterized method \cite{Blinne:2013via}, replacing the kinetic momentum ${\mathbf p}$ with the canonical momentum ${\mathbf q}$ via $ {\mathbf q} - e {\mathbf A} (t)$, and the partial differential equation(PDE) system for the 16 Wigner components can be reduced to ten ordinary differential equations(ODEs). And the nonvanishing Wigner coefficients are:
 \begin{equation}
{\mathbbm w} = ( {\mathbbm s},{\mathbbm v}_i,{\mathbbm a}_i,{\mathbbm t}_i)
\, , \quad  {\mathbbm t}_i := {\mathbbm t}_{0i} -   {\mathbbm t}_{i0}  \, .
\end{equation}
 Because due to the Wigner coeffecients equations of motions are quite lengthy thus here we refrain
 repeating the respective formula form. For the detailed derivations and explicit form we refer the
reader to \cite{Hebenstreit:2011pm,Kohlfurst:2015zxi}.
The corresponding vacuum nonvanishing initial values are
\begin{equation}
{\mathbbm s}_{vac} = \frac{-2m}{\sqrt{{\mathbf p}^2+m^2}} \, ,
\quad  {\mathbbm v}_{i,vac} = \frac{-2{ p_i} }{\sqrt{{\mathbf p}^2+m^2}} \, .
\end{equation}
In the following, one can expresses the scalar Wigner coefficient
by the one-particle momentum distribution function
\begin{equation}
f({\mathbf q},t) = \frac 1 {2 \Omega(\mathbf{q},t)} (\varepsilon - \varepsilon_{vac} ).
\end{equation}
where $\Omega(\mathbf{q},t)= \sqrt{{\mathbf p}^2(t)+m^2}=
\sqrt{m^{2}+(\mathbf{q}-e\mathbf{A}(t))^{2}}$ is the total energy of the electron's (positron's) and
$\varepsilon = m {\mathbbm s} + p_i {\mathbbm v}_i$
is the phase-space energy density.
In order to precisely calculate to one-particle momentum distribution function $f({\mathbf q},t)$, referring to \cite{Blinne:2013via}, it is helpful to introduce an auxiliary three-dimensional vector $\mathbf{v}(\mathbf{q},t)$:
\begin{equation}
v_i (\mathbf{q},t) : = {\mathbbm v}_i (\mathbf{p}(t),t) -
(1-f({\mathbf q},t))  {\mathbbm v}_{i,vac} (\mathbf{p}(t),t) \, .
\end{equation}
 So the one-particle momentum distribution function $f({\mathbf q},t)$ can be obtained by solving the following ordinary differential equations,
\begin{equation}
\begin{array}{l}
\dot{f}=\frac{e\mathbf{E}\cdot \mathbf{v}}{2\Omega},\\
\dot{\mathbf{v}}=\frac{2}{\Omega^{3}}[(e\mathbf{E}\cdot \mathbf{p})\mathbf{p}-e\mathbf{E}\Omega^{2}](f-1)-\frac{(e\mathbf{E}\cdot \mathbf{v})\mathbf{p}}{\Omega^{2}}-2\mathbf{p}\times \mathbbm{a}-2m\mathbbm{t},\\
\dot{\mathbbm{a}}=-2\mathbf{p}\times \mathbf{v},\\
\dot{\mathbbm{t}}=\frac{2}{m}[m^{2}\mathbf{v}-(\mathbf{p}\cdot \mathbf{v})\mathbf{p}],
\end{array}\label{eq3}
\end{equation}
with the initial conditions $f(\mathbf{q},-\infty)=0$, $\mathbf{v}(\mathbf{q},-\infty)=\mathbbm{a}(\mathbf{q},-\infty)=\mathbbm{t}(\mathbf{q},-\infty)=0$, where the time derivative is indicated by a dot, $\mathbbm{a}(\mathbf{q},t)$ and $\mathbbm{t}(\mathbf{q},t)$ are the three-dimensional vectors corresponding to Wigner components, and
 $\mathbf{A}(t)$ denotes the vector potential of the external field.

Finally, by integrating the distribution function $f(\mathbf{q},t)$ over full momentum space, we obtain the number density of created pairs defined at asymptotic times $t\rightarrow+\infty$:
\begin{equation}\label{14}
  n = \lim_{t\to +\infty}\int\frac{d^{3}q}{(2\pi)^ 3}f(\mathbf{q},t) \, .
\end{equation}

For the sake of keeping a self-sustained, let us turn to introduce the semiclassical approximation for spinor QED.
The effects of the field polarization on momentum distribution, especially the interference pattern can be qualitatively and quantitatively interpreted within a semiclassical analysis by means of an effective scattering potential based on the WKB approximation \cite{Dumlu1}.
The approximated production number for the created fermionic particles takes the form (cf.\cite{Dumlu1})
\begin{equation}\label{11}
 N^{spinor}_{q}\approx \sum_{t_{p}}e^{-2K^{(p)}_{\mathbf{q}}}+\sum_{t_{p}\neq t_{p^{'}}}2\cos(2\alpha^{(p,p^{'})}_{\mathbf{q}})(-1)^{p-p^{'}}e^{-K^{(p)}_{\mathbf{q}}-K^{(p^{'})}_{\mathbf{q}}},
\end{equation}
where $t_{p}$ are the turning points by $\Omega(\mathbf{q},t_{p})=\sqrt{m^{2}+[\mathbf{q}-e\mathbf{A}(t)]^{2}}=0$, $K^{(p)}_{\mathbf{q}} =  \left| \int^{t_{p}}_{t^{*}_{p}} \Omega(\mathbf{q},t)dt \right|$, and $\alpha^{(p,p^{'})}_{\mathbf{q}}=  \int^{t_{p^{'}}}_{t_{p}} \Omega(\mathbf{q},t)dt $ is the phase accumulation related to the different turning point pairs, for a more detailed discussion, see Ref \cite{Dumlu1}. As was discussed in \cite{Dumlu1}, the first term is described as every distinct pair of turning points for pair production, whereas the second one is interference term of different turning points which is responsible for the oscillation in the momentum spectrum.

For example, if a single pair of turning point dominates, then there is no interference pattern in momentum distribution. Therefor the WKB result for the created number of pairs for momentum $\mathbf q$ is given by
\begin{equation}\label{4}
  N^{spinor}_{q}\approx \exp(-2K) \, , \quad
 K =  \left| \int^{t_{2}}_{t_{1}}  dt \Omega(\mathbf{q},t)dt \right| \, ,
\end{equation}
where $t_{1}$ and $t_{2}$ are dominant turning points closed to the real $t$ axis. This is the case that was studied in Ref. \cite{Akkermans:2011yn}. For two pairs of complex turning points, the production rate was then estimated to be a sum of two terms which takes the form
\begin{equation}\label{5}
  N^{spinor}_{q}\approx e^{-2K_{1}}+e^{-2K_{2}}-2\cos(2\alpha)e^{-K_{1}-K_{2}},
\end{equation}
which is then explain interference effects in the momentum spectra.

\section{Momentum spectra}\label{result1}

\begin{figure}[h]\suppressfloats
\includegraphics[width=15cm]{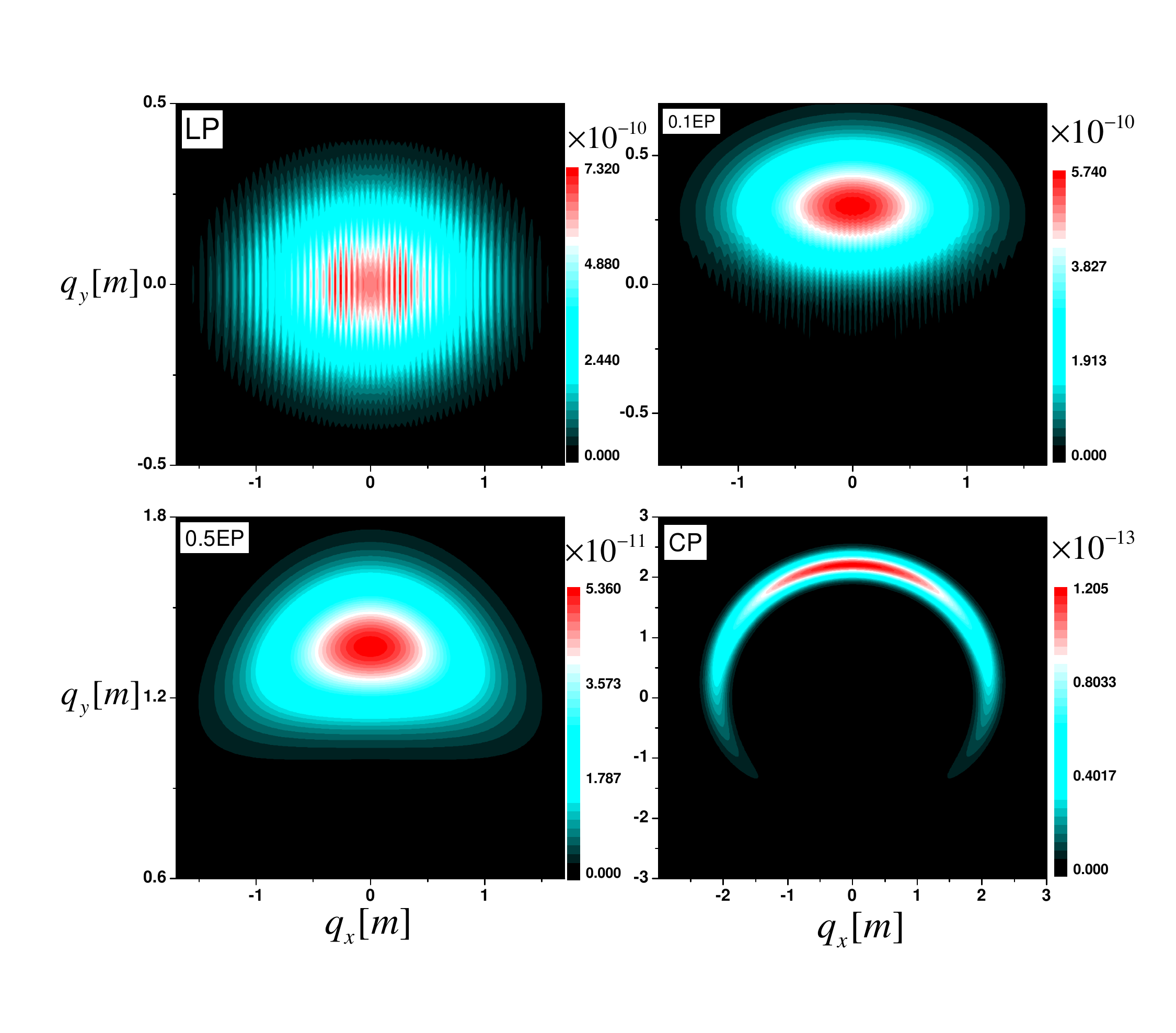}
\caption{Momentum spectra of created $e^{-}e^{+}$ pairs in the($q_{x},q_{y}$)plane(where($q_{z}=0$)). These plots are for the strong field $E_{1s}(t)$ with different polarizations. From top left to bottom right the values of polarization parameters are $\delta=0, 0.1, 0.5, 1$, respectively. The other field parameters are $E_{1s}=0.1\sqrt{2}E_{cr}$, $\omega=0.05m$, and $\tau=100/m$.}
\label{fig:1}
\end{figure}

By using the DHW method, we compute the momentum spectrum of the produced particles for several values of the high harmonics $b\omega$ when the carrier phase is chosen as $\varphi=0$. Note that the momentum spectra of the created pairs is highly sensitive to the change of carrier phase $\varphi$ \cite{Hebenstreit3,Abdukerim}, however, it is not our main concerned issue in present paper. Instead our main concerns are for typical polarizations as the linear $\delta=0$, the near-linear $\delta=0.1$, the middle elliptic $\delta=0.5$ and the circular $\delta=1$.

\subsection{Single strong field}

\begin{figure}
\centering
  \begin{minipage}{7cm}
   \includegraphics[width=7cm]{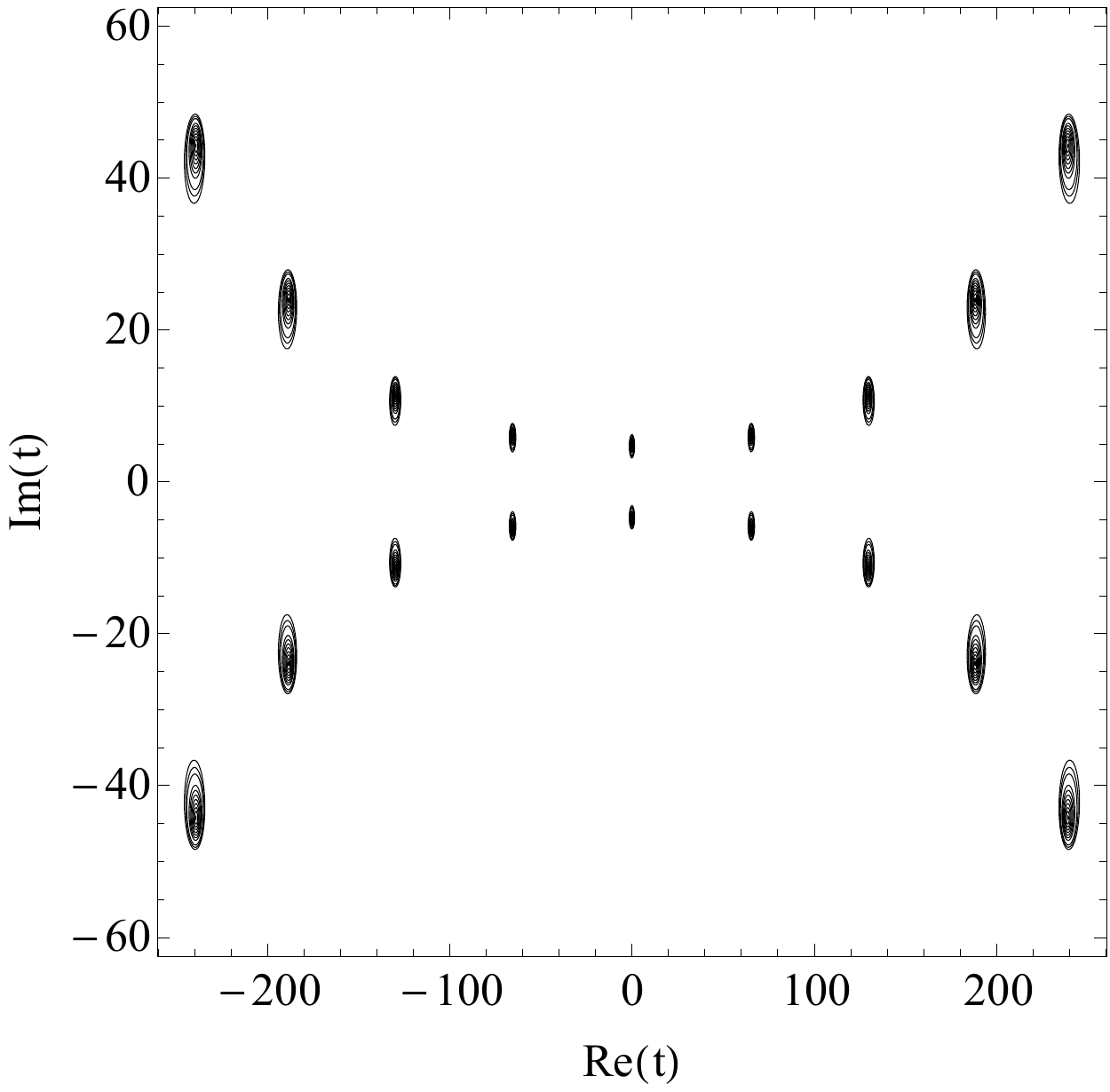}
  \end{minipage}
  \begin{minipage}{7cm}
    \includegraphics[width=7cm]{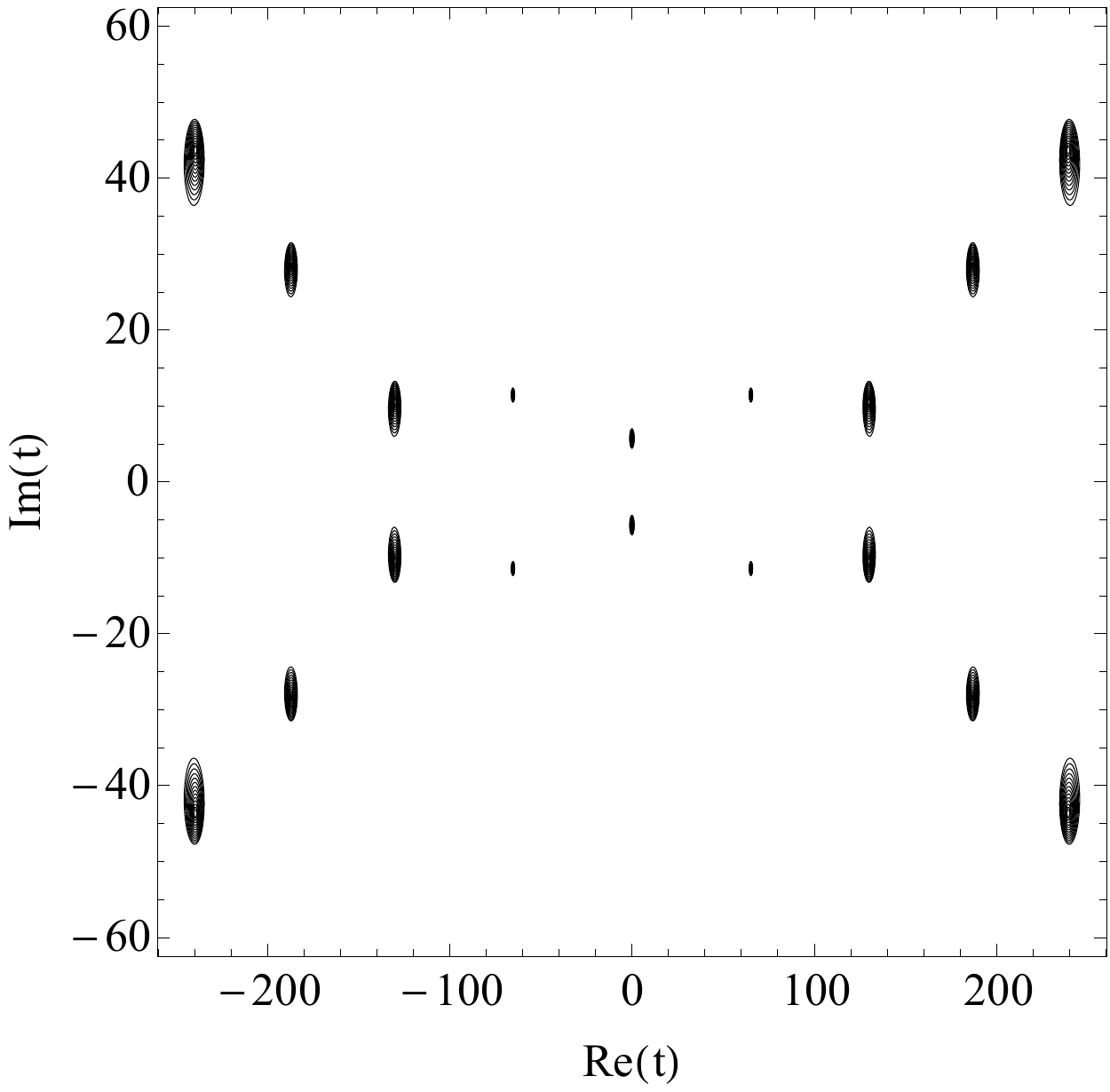}
  \end{minipage}

  \begin{minipage}{7cm}
   \includegraphics[width=7cm]{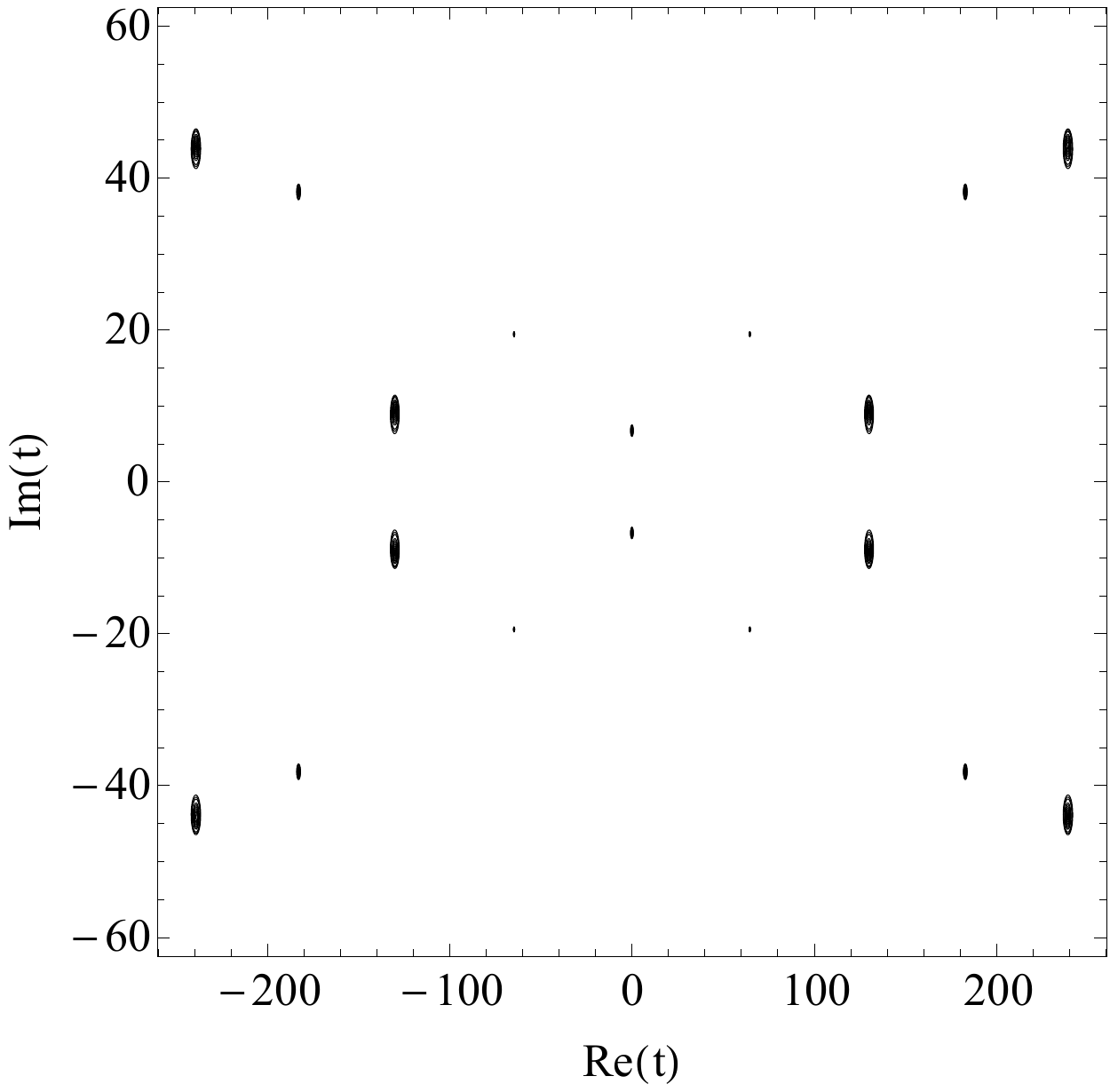}
  \end{minipage}
  \begin{minipage}{7cm}
   \includegraphics[width=7cm]{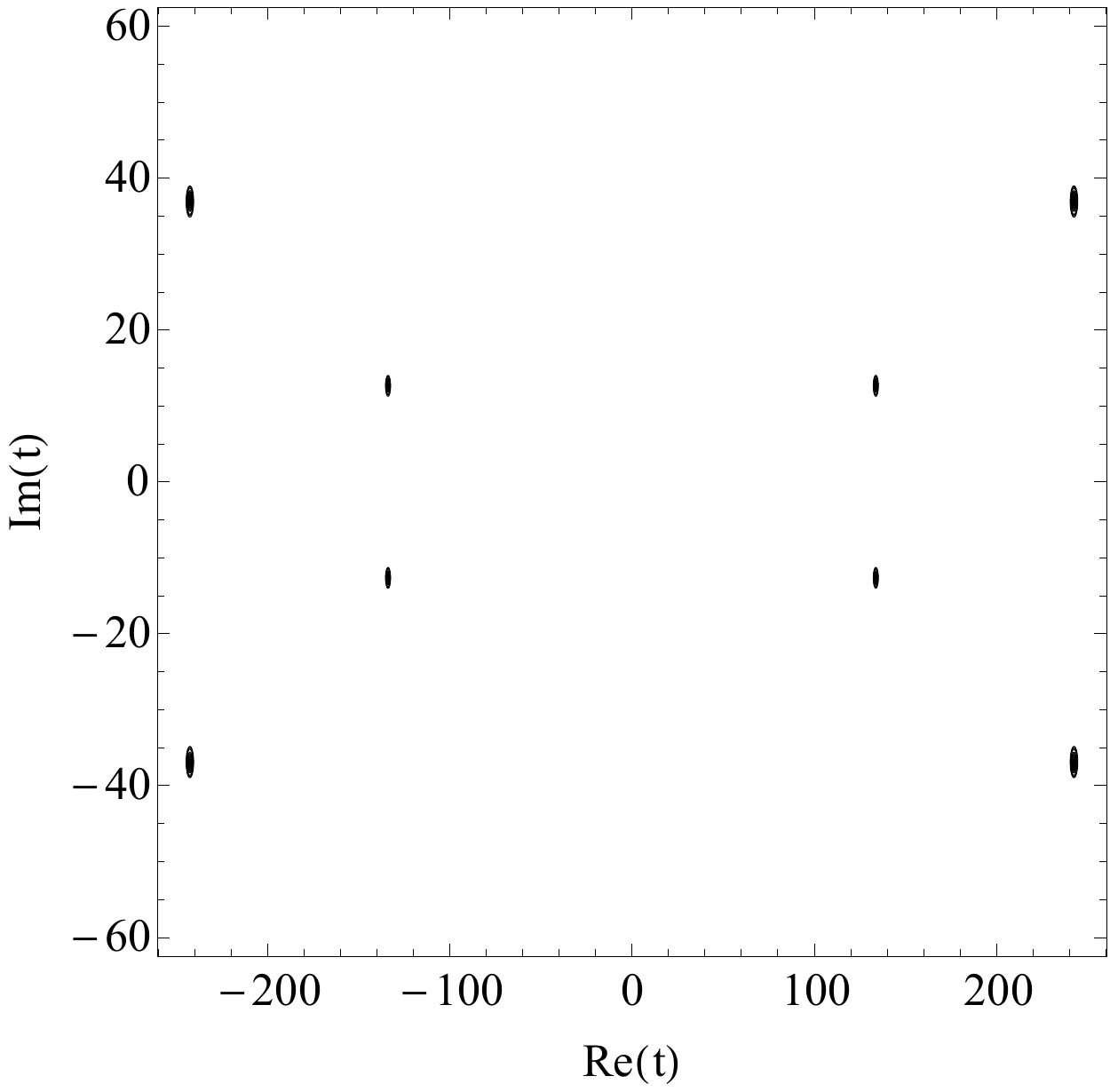}
  \end{minipage}
  \caption{Contour plots of $|\Omega (\mathbf{q},t)|^{2}$ in the complex $t$ plane,
  showing the location of turning points where $\Omega (\mathbf{q},t)=0$.
   These plots are for the strong field $E_{1s}(t)$  with different polarizations. The other field parameters are
  $E_{1s}=0.1\sqrt{2}E_{cr}$, $\omega=0.05m$, and $\tau=100/m$.
  From top left to bottom right the values of polarization parameters are $\delta=0, 0.3, 0.5, 1$,
  respectively,
  and the momentum values (in units of $m$) are $(q_{x}=0,q_{y}=0) , (q_{x}=0,q_{y}=0.6), (q_{x}=0,q_{y}=1.35), (q_{x}=0,q_{y}=2.0)$,
   respectively.}
  \label{S}
\end{figure}

By solving Eq.~\eqref{eq3} numerically, we can obtain the created particles distribution functions. Fig. \ref{fig:1} depicts the two-dimensional momentum distribution in the $(q_{x},q_{y})$ plane for the weak frequency strong field $\mathbf{E}_{1s}(t)$ when $\mathbf{E}_{2w}(t)=0$. From Fig. \ref{fig:1} (LP case), one can see a strong oscillation characteristic appeared in spectra for the linearly polarized case. And the peak position of distribution is also located at the center of momentum spectrum, which agrees with results of Refs. \cite{Li:2015cea,Hebenstreit3,Dumlu:2010vv}. For the nonzero polarization, however, the momentum spectrum looses its oscillation structure gradually and its peak shifts along the positive $q_{y}$ direction with a decreasing of peak value when polarization parameter $\delta$ increases until to that a ringlike shape appearing in spectrum for $\delta\sim1$, see Fig. \ref{fig:1} (CP).
To understand that the momentum spectrum are not peaked around $q_{x/y}=0$ but shift along the $q_y$ axis for the $\delta \neq 0$ is because mostly pairs created in the $E_{x}(t)$ component of the electric field has to be accelerated or/and decelerated along the $y$ direction by the $E_{y}(t)$ component. Due to the even and odd function properties of $x$ and $y$ field components, thus, the momentum spectrum exhibit the symmetric/asymmetric behavior in $q_x/q_y$ directions.

For the given electric field Eq.~\eqref{eq1}, there exists an infinite number of turning points, and
the related dominant turning points are numerically obtained when the effective scattering potential $\Omega(\mathbf{q},t_{p})=\sqrt{m^{2}+[\mathbf{q}-e\mathbf{A}(t)]^{2}}$ is zero for certain $\mathbf{q}$. From the Fig. \ref{fig:1}, one can infer that, for nonzero polarization parameters, the oscillation or interference in momentum spectrum vanish. Because pair production in a strong field is a non-Markovian poroses \cite{Schmidt}, the cosine function in equation~\eqref{11} has a time integral which consequently causes the accumulated phases $\alpha^{(p,p^{'})}_{\mathbf{q}}=  \int^{t_{p^{'}}}_{t_{p}} \Omega(\mathbf{q},t)dt $ to depend on the complete earlier history; so even the small changes of the electric field parameters can easily change relative phases of the amplitude naturally. Thus we can understand that the variation of the field polarization can easily change dominant turning points location in the complex time plane.

To clarify the above discussion, we plot the location of the complex conjugate pair of turning points in the complex time $t$ plane, which is shown in Fig. \ref{S}. When $\delta=0$, the location of pairs of turning points agrees with the result \cite{Dumlu:2010vv}. As the field polarization increases, the positions of dominating pair of turning points close to the real $t$ axis also change. Obviously turning points departs from the real time $t$ axis when one goes from linear to circular polarization. Therefore, the interference effects between those pairs of turning points become weaker and weaker and even vanishing with field polarization increases, which results in vanishing of oscillatory structure of the spectrum. In Fig. \ref{S}, it is evident that the location of turning points in the case of circular polarization, see Fig. \ref{fig:1} (CP), are far away from real axis and also the fewer number of points exist compare to other cases.
Consequently, the interference between these turning points become weaker as the increasing filed polarization for the few-cycle electric field in Eq.~\eqref{eq1}, which is only appropriate for the weak frequency but strong field, {\it i.e.}, for the $\mathbf{E}_{1s}(t)$. \footnote{Note that in the multiphoton scenario as studied in \cite{olugh}, such a weak oscillatory pattern is present in spectra when one goes from liner to circular polarization without chirp. Moreover, note that $E_{y}(t)$ component of the field is an odd function of $t$, and the $A_{y}(t)$ is thus an even function, resulting the created particles momentum distribution being asymmetric about $q_{y}$ axes.}

As a comparison study, let us now turn to the results of the momentum spectrum for dynamically assisted two-color field for different polarization and different high harmonics.

\subsection{Two-color field when $\delta=0$ and $\delta=0.1$}

\begin{figure}[h]\suppressfloats
\includegraphics[width=15cm]{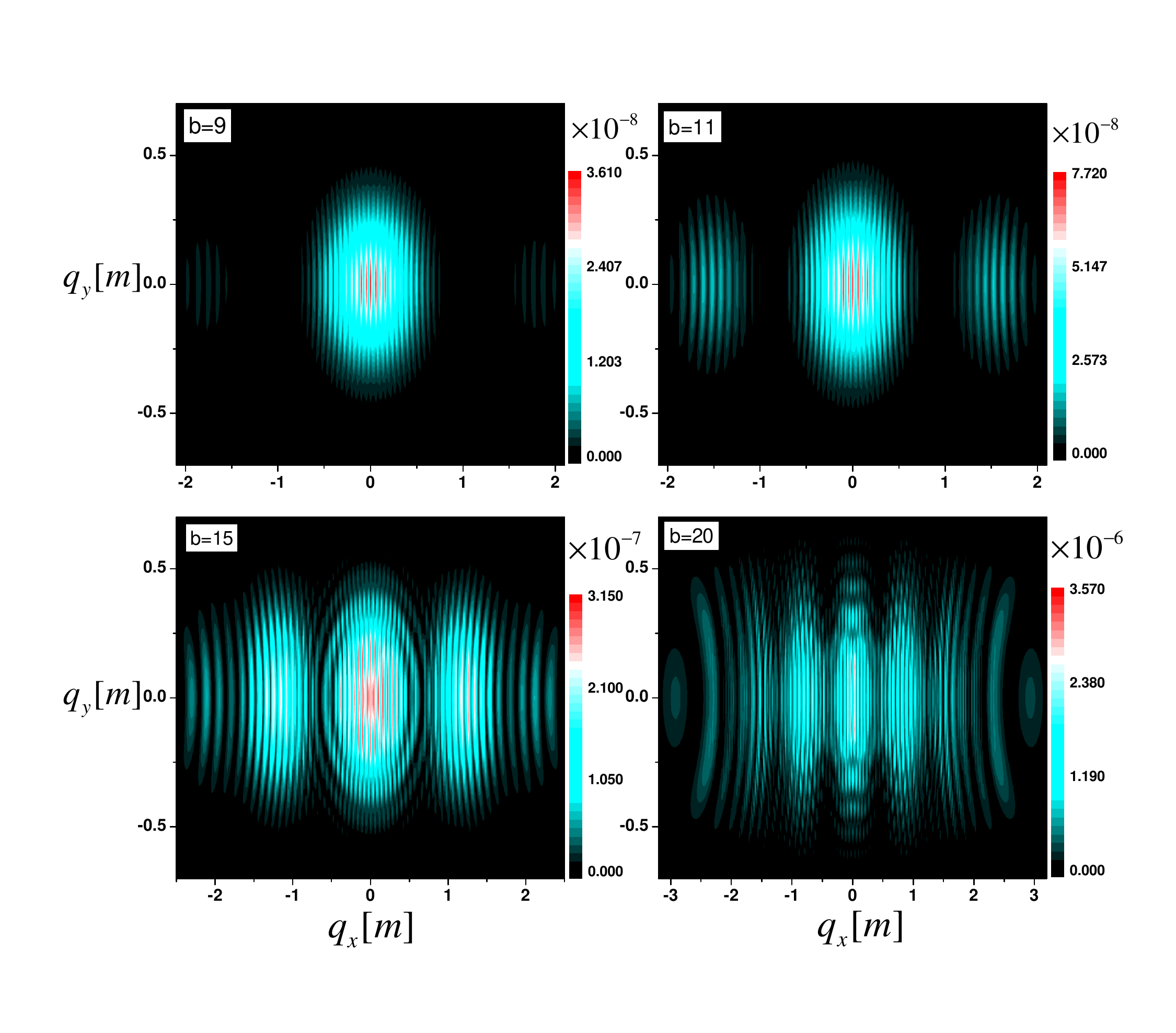}
\caption{Momentum spectra of created $e^{-}e^{+}$ pairs in the ($q_{x},q_{y}$) plane(where($q_{z}=0$)). These plots are for linear polarized ($\delta=0$) combined field $E(t)$. The high harmonics $b\omega=0.45m, 0.55m, 0.75m, 1m$ from top left to bottom right, respectively. And the other field parameters are $E_{1s}=0.1\sqrt{2}E_{cr}$, $E_{2w}=0.01\sqrt{2}E_{cr}$, $\omega=0.05m$, and $\tau=100/m$.}
\label{fig:2}
\end{figure}

The particle momentum spectra under dynamically assisted field is shown in Fig. \ref{fig:2} for linear polarization case $\delta=0$. One can see that the momentum spectrum exhibits strong and rapid oscillation structure. Obviously, there exist some discrete side maxima at the large momentum regime. We think it is due to the impact of the high-frequency but weak field with high harmonics $b\omega$ superimposed to the strong field. To our knowledge, in the presence of a non-perturbative field $E_{1s}$, the produced pairs are continuously accelerated, and created particle momentum is mainly determined by its creation time. At the earlier time it is created then it has to be accelerated at the longer time and finally it gets the higher longitudinal momentum. This prediction is also supported by the Refs.~\cite{Orthaber:2011cm,Sitiwaldi:2018wad}. As high harmonics $b\omega$ increases it is more remarkable. Meanwhile the peak values of momentum distribution are increased to $2$ and $4$ orders larger compared to linearly polarized case of a single strong field, respectively, when $b\omega\approx 0.5m$ and $b\omega\approx 1m$. These amplification results mainly attribute to the dynamically assisted Sauter-Schwinger effect.

For the interference effects of created particles in spectra is again due to the interaction between the complex conjugate pairs of turning points. These dominate pairs of turning points distribution gradually approaches real $t$ axis more and more with $b\omega$ and finally the different sets of turning pontes are almost equidistant from real $t$ axis. This results in the complicated and pronounced oscillation structures of the momentum spectra.

\begin{figure}[h]\suppressfloats
\includegraphics[width=15cm]{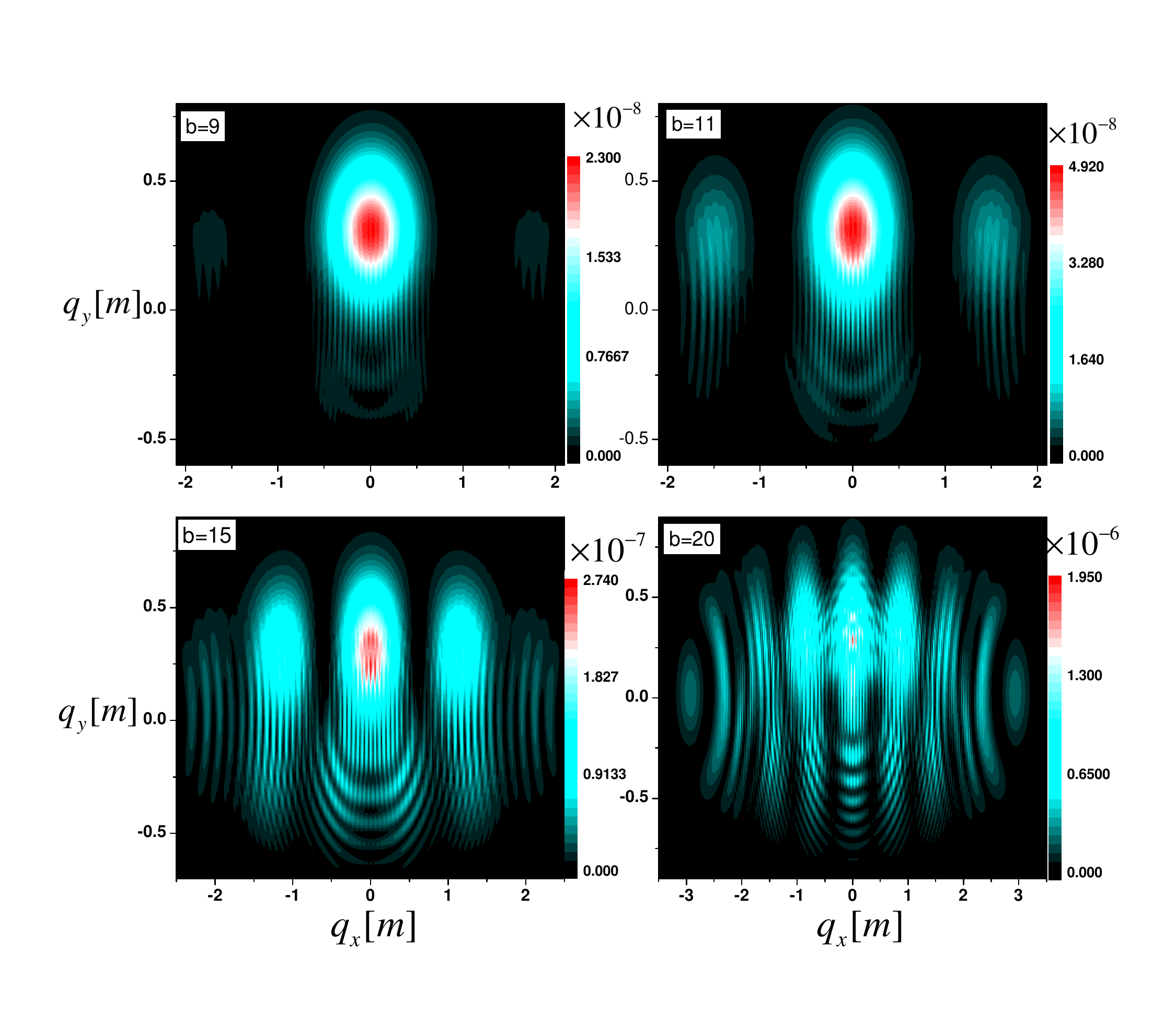}
\caption{Momentum spectra of created $e^{-}e^{+}$ pairs in the ($P_{x},P_{y}$) plane(where($P_{z}=0$)). These plots are for the near-linear elliptically polarized ($\delta=0.1$) combined field $E(t)$. The high harmonics $b\omega=0.45m, 0.55m, 0.75m, 1m$ from top left to bottom right, respectively. And the other field parameters are $E_{1s}=0.1\sqrt{2}E_{cr}$, $E_{2w}=0.01\sqrt{2}E_{cr}$, $\omega=0.05m$, and $\tau=100/m$.}
\label{fig:3}
\end{figure}

For the near-linearized case $\delta=0.1$, the momentum spectrum is exhibited in Fig. \ref{fig:3}. Comparable to the single strong field case, in the case of two-color fields, the peak value of spectrum increases and the range of spectrum expands in momentum space. And this tendency is more striking with $b\omega$. Meanwhile, the interference effects also become stronger with $b\omega$. For example, for small $b\omega$, the interference pattern appear only along the negative $q_{y}$, while for large $b\omega$, the interference effects occur also along the positive $q_{y}$.

\begin{figure}[h]\suppressfloats
\includegraphics[width=15cm]{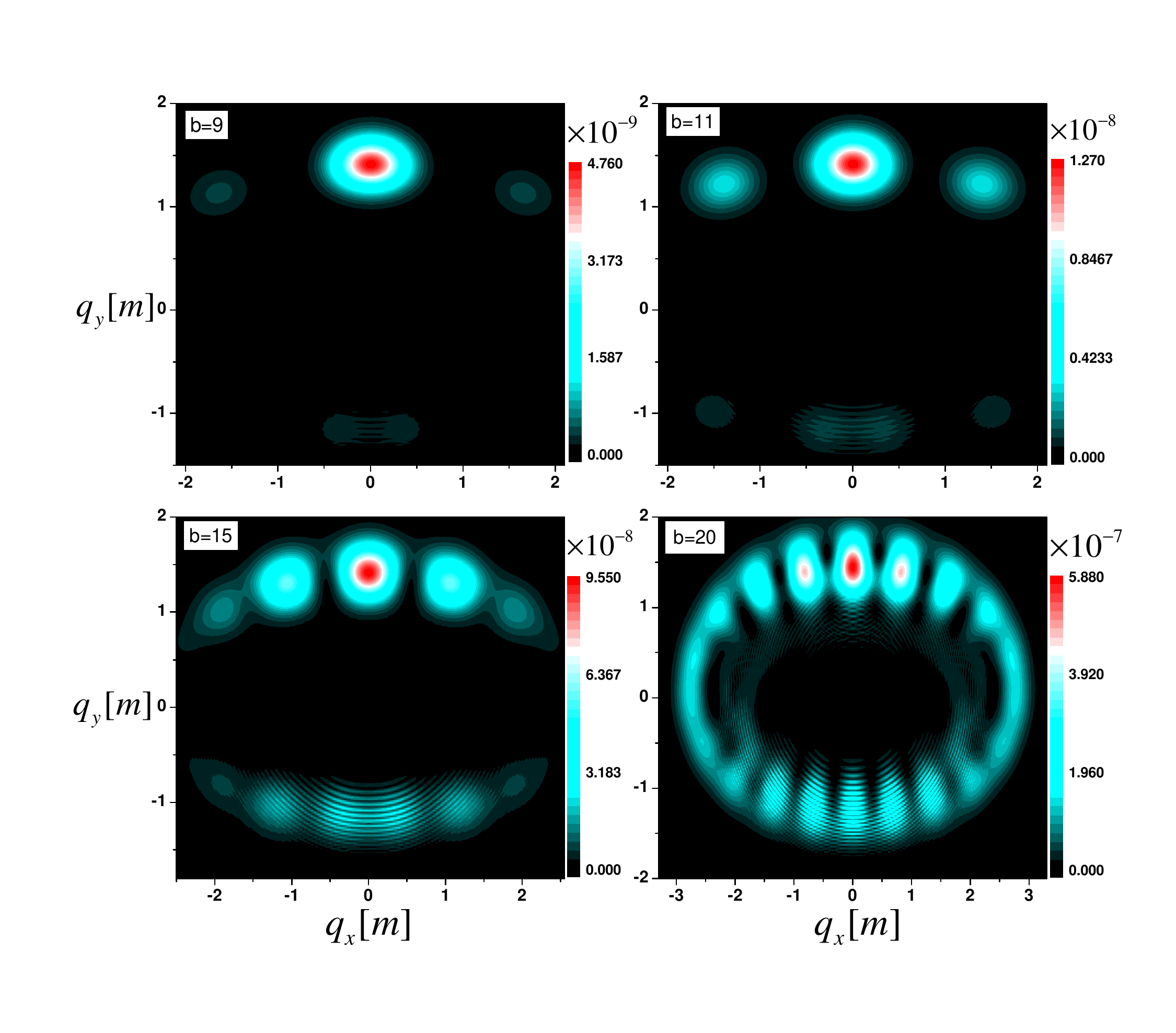}
\caption{Momentum spectra of created $e^{-}e^{+}$ pairs in the ($P_{x},P_{y}$) plane(where($P_{z}=0$)). These plots are for the middle elliptically polarized ($\delta=0.5$) combined field $E(t)$. The high harmonics $b\omega=0.45m, 0.55m, 0.75m, 1m$ from top left to bottom right, respectively. And the other field parameters are $E_{1s}=0.1\sqrt{2}E_{cr}$, $E_{2w}=0.01\sqrt{2}E_{cr}$, $\omega=0.05m$, and $\tau=100/m$.}
\label{fig:4}
\end{figure}

\subsection{Two-color field when $\delta=0.5$}

For middle-elliptical polarization case $\delta=0.5$, the result of momentum spectrum is shown in Fig. \ref{fig:4}. From the top left of Fig.\ref{fig:4}, when $b\omega=0.45m$, one can see that the $e^{-}e^{+}$ pairs locates mainly in the regime of positive $q_y$ while a weak interference effect appears in the regime of negative $q_{y}$. With the increase of $b\omega$, we can observe multiple peaks appear at the same time. On the other hand, stronger interference effects at the negative $q_{y}$ plane occur, see the lower panel of Fig. \ref{fig:4}. When $b\omega=1m$, the spectrum has the $14$ peaks.

These results are very similar to the findings of the Ref. \cite{Otto}. These multiple peaks are referred to as shell structures. In general the peak pattern is dominated by the strong but slowly varying field $E_{1s}$, while the second weak but rapidly changing field $E_{2w}$ is mainly responsible for the appearances of additional peaks, which are not visible in the case of a single strong field $E_{1s}$ alone. Therefore the observed multiple peaks form a shell structure exhibiting a lifting pattern when the $b\omega$ is large and the middle-ellipticity is applied which destroy the symmetry of momentum distribution along $q_y$.

In particular, when $b\omega=1m$, a pearl-necklace-like pattern is observed in the spectrum with more pronounced interference pattern. The interference effects are even evident at the positive $q_{y}$. For this case, there would be more complex conjugate turning points pairs having approximately the same distance to the real $t$ axis. Thus, the distinct interference effects occur in the momentum spectrum.

By comparing with the single strong field of middle-elliptical polarization shown in Fig.\ref{fig:1}(0.5EP), we find that, in the case of two-color field, the momentum spectrum becomes larger with $b\omega$. Finally, the peak values of spectra are enhanced from $5.36\times10^{-11}$ (for single strong field $E_{s}$) to $5.88\times10^{-7}$ (two-color field when $b\omega=1m$). Note that in this case the position of spectrum peak often appears at the positive $q_{y}$ plane and $q_{x}=0$.

\begin{figure}[h]\suppressfloats
\includegraphics[width=15cm]{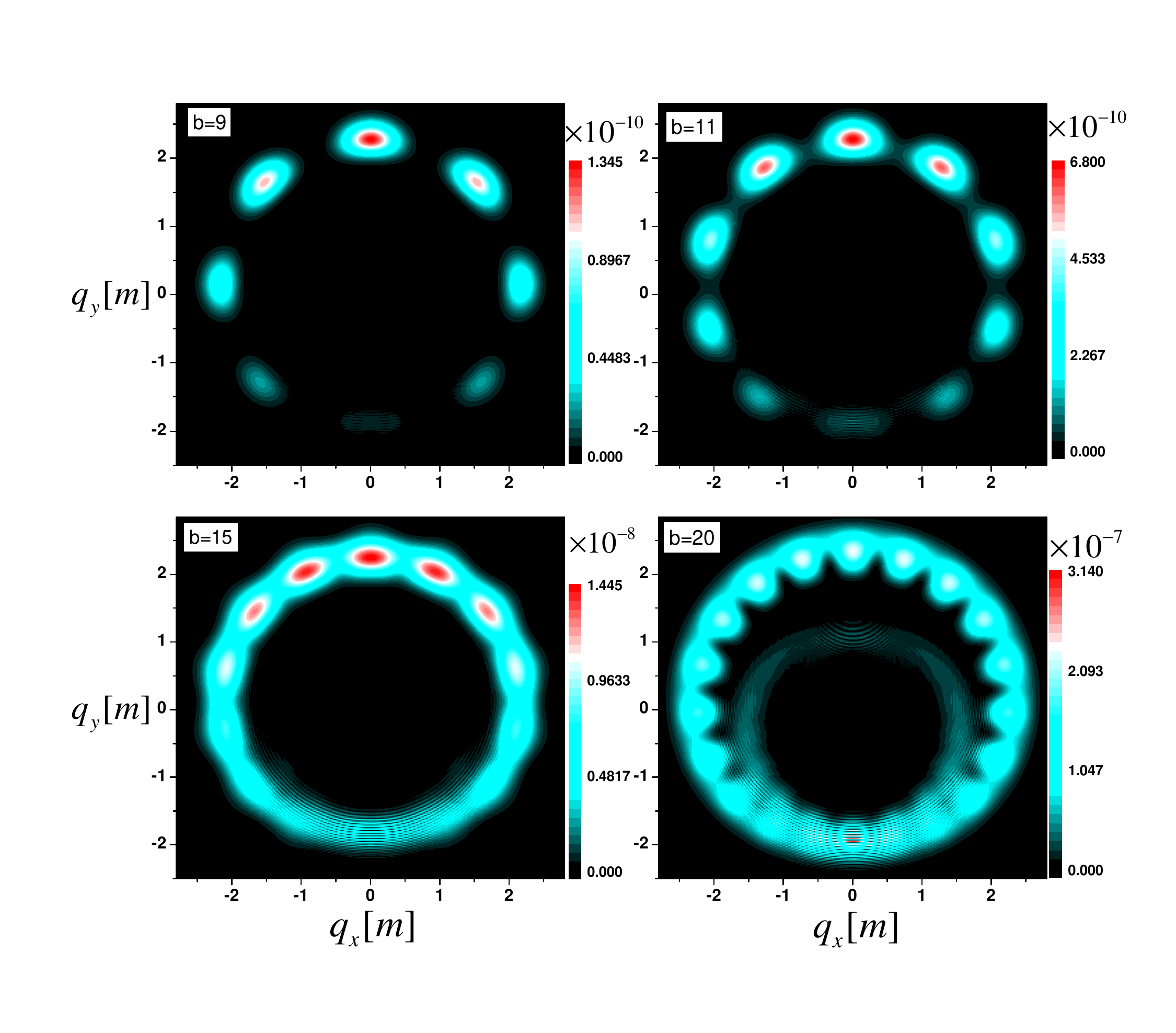}
\caption{Momentum spectra of created $e^{-}e^{+}$ pairs in the ($P_{x},P_{y}$) plane(where($P_{z}=0$)). These plots are for the circular polarized ($\delta=1$) combined field $E(t)$. The high harmonics $b\omega=0.45m, 0.55m, 0.75m, 1m$ from top left to bottom right, respectively. And the other field parameters are $E_{1s}=0.1\sqrt{2}E_{cr}$, $E_{2w}=0.01\sqrt{2}E_{cr}$, $\omega=0.05m$, and $\tau=100/m$.}
\label{fig:5}
\end{figure}

\begin{figure}
\centering
  \begin{minipage}{7cm}
   \includegraphics[width=7cm]{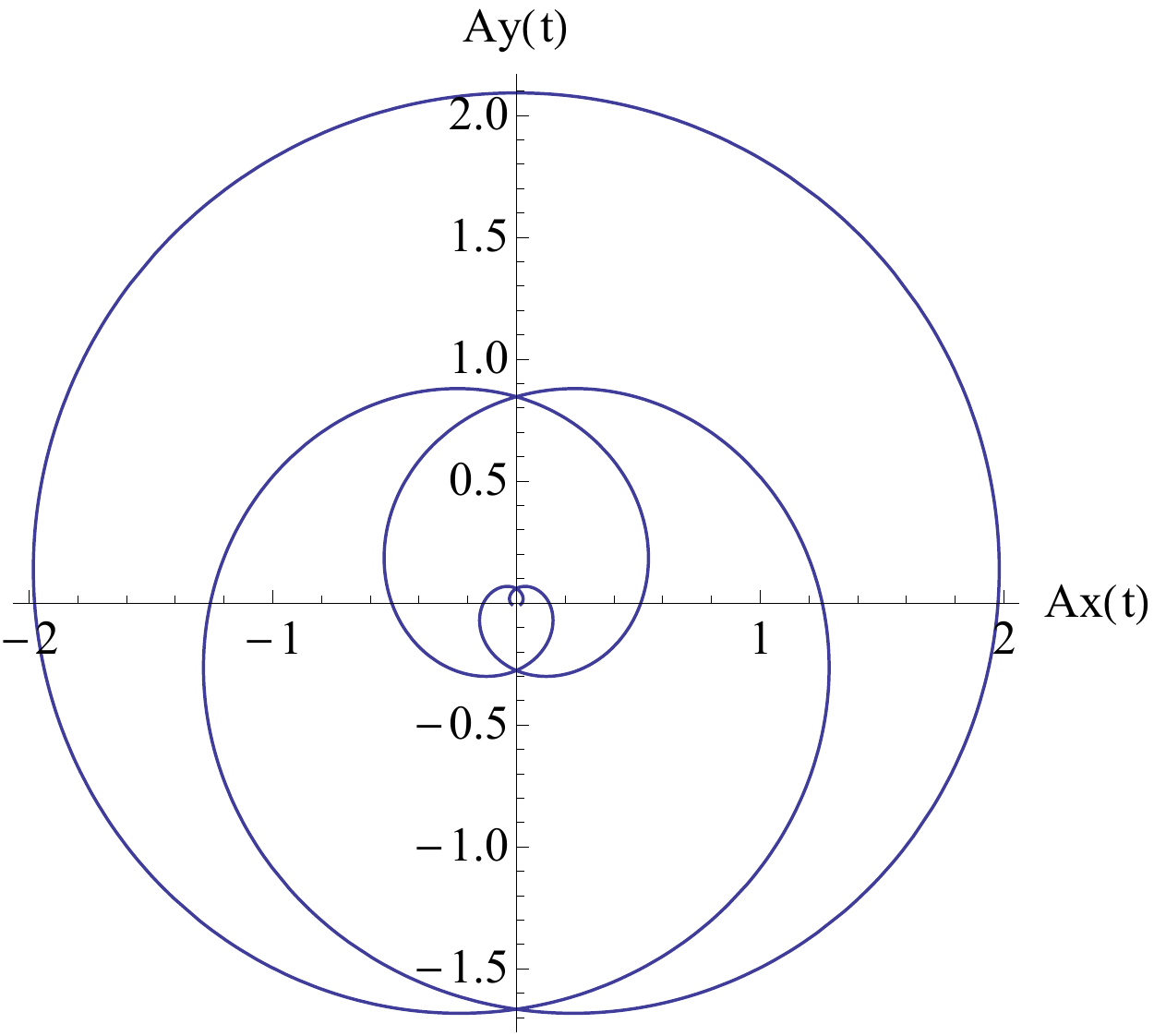}
  \end{minipage}
  \begin{minipage}{7cm}
    \includegraphics[width=7cm]{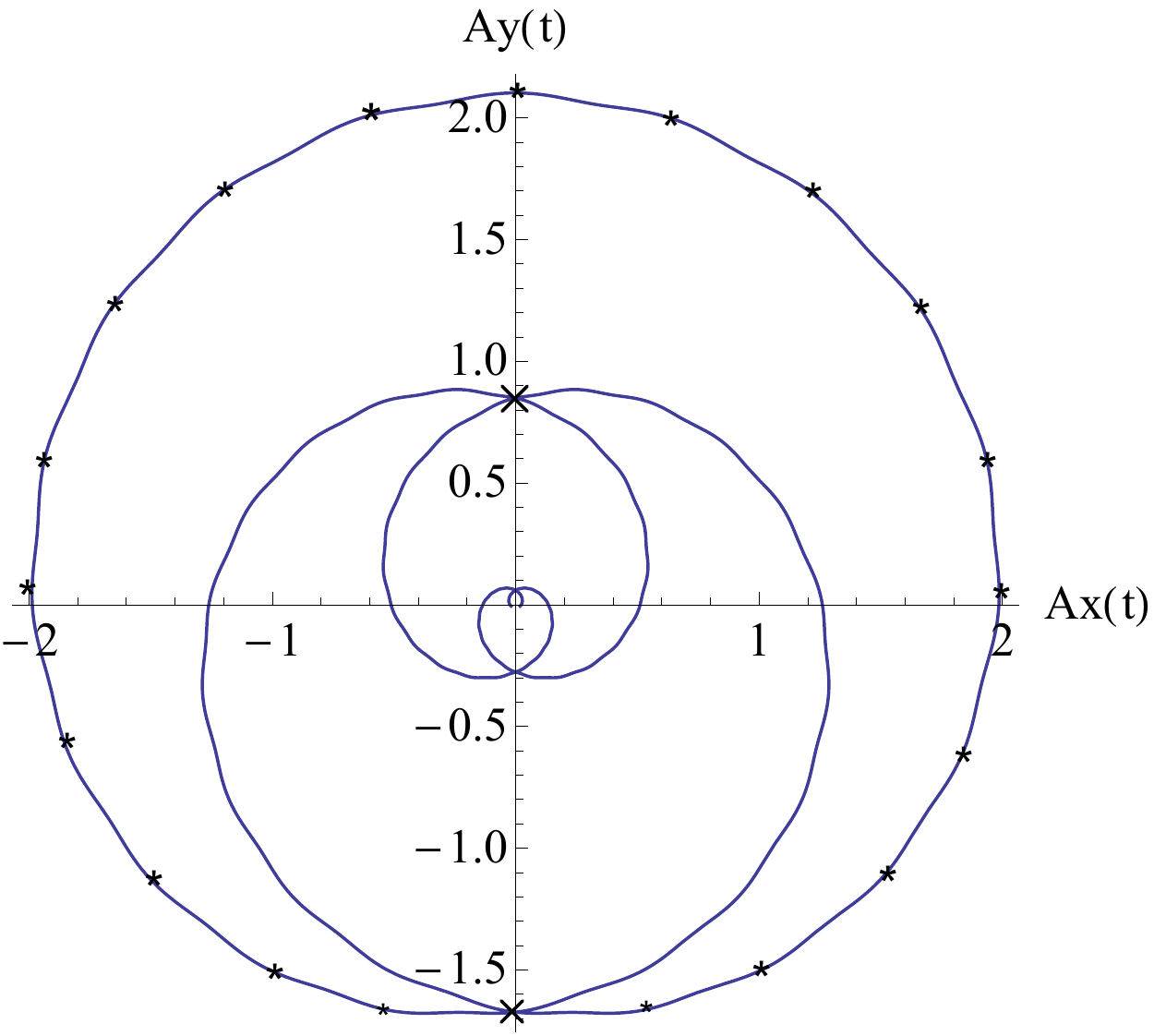}
  \end{minipage}
  \caption{The corresponding vector potential $A(t)$ for time dependence of the electric field $E(t)$ in units of the critical
field for the circularly polarized ($\delta=1$) case.
Left panel is vector potential for strong field $E_{1s}(t)$ without superimposed weak field. This right displays vector potential for the dynamically assisted field $E(t)$ with a high harmonics $b\omega=1m$. The black stars give the predicted
pair production peaks as given by Eq. \eqref{q}. The chosen parameters are $E_{1s}=0.1\sqrt{2}E_{cr}$, $E_{2w}=0.01\sqrt{2}E_{cr}$, $\omega=0.05m$, and $\tau=100/m$
where $m$ is the electron mass.}
  \label{fig:A(t)}
\end{figure}

\begin{figure}
\centering
  \begin{minipage}{7cm}
   \includegraphics[width=7cm]{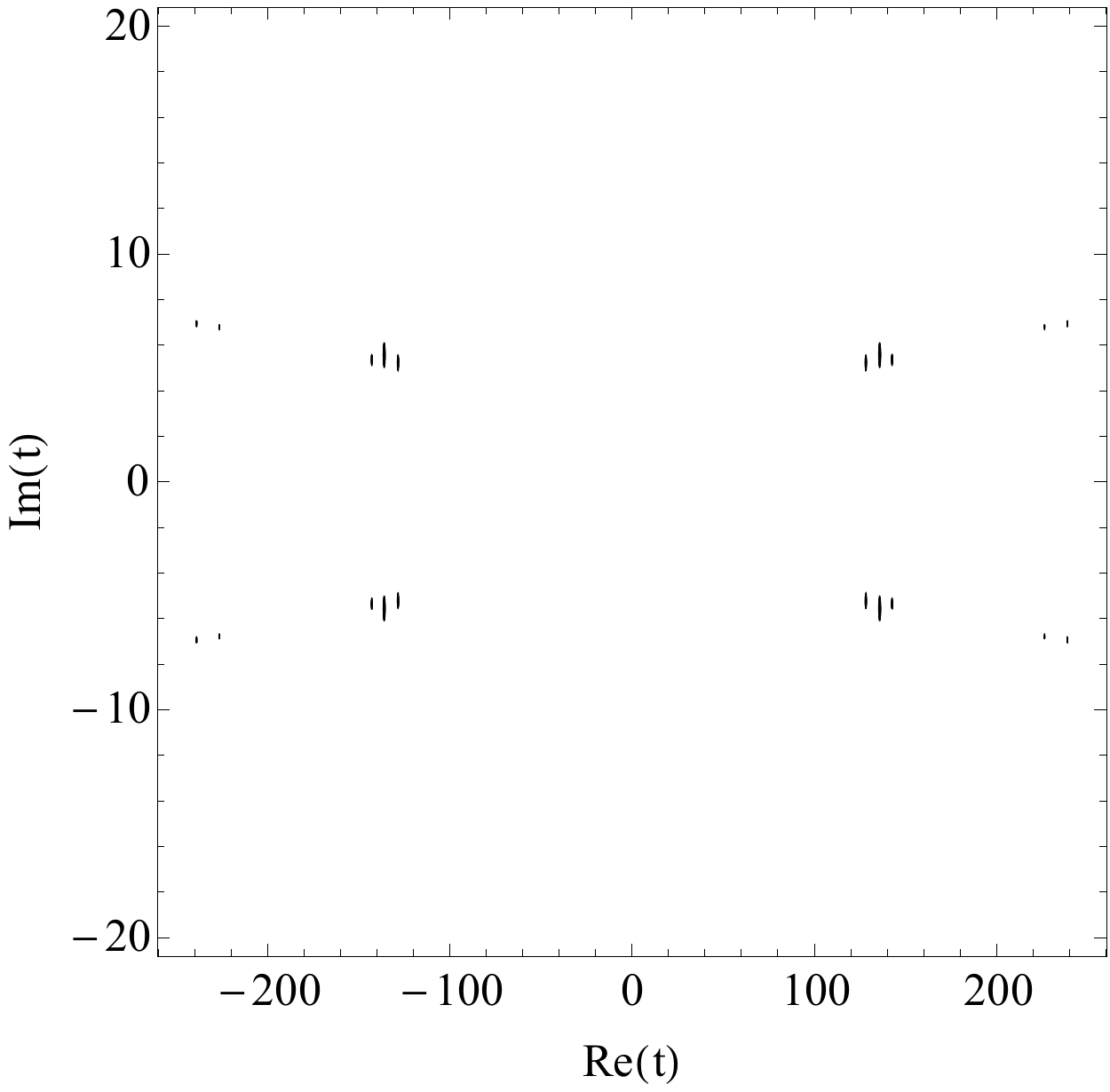}
  \end{minipage}
  \begin{minipage}{7cm}
    \includegraphics[width=7cm]{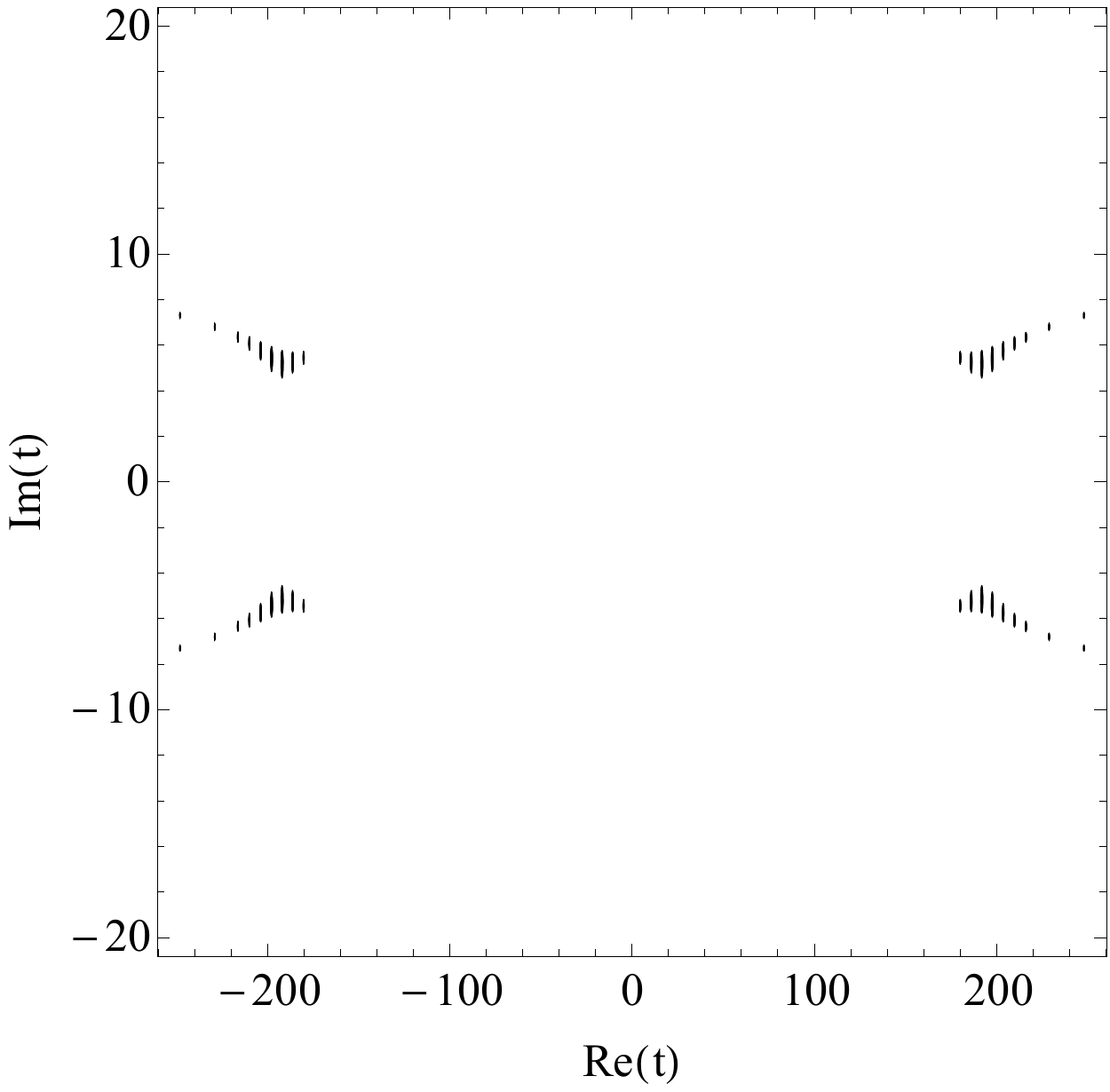}
  \end{minipage}
  \caption{Contour plots of $|\Omega (\mathbf{q},t)|^{2}$ in the complex $t$ plane,
  showing the location of turning points where $\Omega (\mathbf{q},t)=0$.
  These plots are for circular polarized ($\delta=1$) combined field $E(t)$ when high harmonics $b\omega=1m$. The other field parameters are $E_{1s}=0.1\sqrt{2}E_{cr}$, $E_{2w}=0.01\sqrt{2}E_{cr}$, $\omega=0.05m$, and $\tau=100/m$.
  From left to right the momentum values (in units of $m$) are $(q_{x}=0,q_{y}=2.25), (q_{x}=0,q_{y}=-1.90)$,
   respectively.}
  \label{CP}
\end{figure}

\subsection{Two-color field when $\delta=1$}

For the circular polarized case $\delta=1.0$, the momentum spectra is shown in Fig. \ref{fig:5} for different $b\omega$.
For $b\omega=0.45m$, we can observe that $8$ peaks are centered around the origin with an exhibition of the weak interference pattern at the bottom of the negative $q_{y}$. As $b\omega$ increases, the number of multiple peaks of the momentum spectrum increases also and the separated adjacent peaked ranges of spectra are gradually approaching until to connected with each other. For example, for $b\omega=0.75m$, there appears a complete ring-like form in the momentum spectrum. Meanwhile it is found that there is always $(b-1)$ peaks when $b\omega$ increases until to $0.95m$ with the strong interference signature at the negative $q_{y}$ plane. Moreover in all cases of $b\omega\leq 0.95m$ the maximum of peaks persist at the positive $q_{y}$ regime.

For the circular polarization, if the Gaussian envelop is disregarded, then the homogeneous electric field \eqref{eq1} strength is simplified as
\begin{equation}
\mid\mathbf{E}_{0}(t)\mid=E_{cr}\sqrt{E^{2}_{1s}+E^{2}_{2w}+2E_{1s}E_{2s}\cos[(b-1)\omega t]},
\end{equation}
which has an periodical number of local maxima at $t_{0k}=\frac{2K\pi}{(b-1)\omega}$, $K\in \mathbbm{Z}$. Hereby, $N=\mid b-1\mid$ is the number of maxima for this field. This also holds for the pulsed field where $\mid t_{0k}\mid <\tau$, cf. Ref. \cite{Blinne:2016yzv}. In general, most pairs are expected to appear at those times corresponding to the local maxima of field. Then those produced pairs are subject to acceleration by the electric field, and the gained momenta are 
\begin{equation}\label{q}
\mathbf{q}= \int^{\infty}_{t_{0}}e\mathbf{E}(t)~dt=\mathbf{A}(t_{0})-\mathbf{A}(\infty)=\mathbf{A}(t_{0}).
\end{equation}
It is therefore not surprising, why those $b-1$ multiple peaks are observed in produced pairs spectrum \cite{Blinne:2016yzv}.

An interesting finding is for $b\omega=1m$, strong interference effects are visible at the bottom of the negative $q_{y}$ plane and a sub ring structure appears at the inner part of the multiple peaks ring in the momentum distribution. In contrast to situations $b\omega\leq0.95m$, when $b\omega=1m$, three different aspects are remarkably found: (1) the interference effects at the positive $q_{y}$ plane with the inner ring appears; (2) overall $b=20$ but not $b-1=19$ peaks exist; and (3) the $20^{th}$ peak as the maximum one locates at the negative $q_{y}$ but not the positive $q_{y}$ plane, see Fig.~\ref{fig:5}($b=20$). For understanding these results, it seems that the outer multiple ring and inner sub ring overlaps in the negative $q_{y}$ plane that results in a resonances effect and maybe it leads to an additional peak beside the $b-1$ peaks in the momentum spectrum.

In order to understand the spectra furthermore, we plot the corresponding vector potential $A(t)$ for the circular polarization ($\delta=1$) in Fig.~\ref{fig:A(t)}. On the one hand we find that multiple peaks appears in the
vector potential (right) for the dynamically assisted field $E(t)$ with a high harmonics $b\omega=1m$ as compared to the vector potential (left) for only strong field $E_{1s}(t)$ without superimposed weak field. On the other hand, from the right panel of Fig.~\ref{fig:A(t)} we also find that $19$ (star markers) maximum peak strengths appears in the large outer ring. However, it is found that, at the marked positions by black crosses, the spirales of created pair particle meet at two different times with same final vector potential (momenta). By comparing the right panel of Fig.~\ref{fig:A(t)} with Fig.~\ref{fig:5}($b=20$) we observe that the produced pairs are almost close to the prediction momenta accordance with Eq.~\eqref{q} so that the corresponding interference pattern is observed at the black cross points. Therefor it provides a potential explanation for the number of pair production peaks in outer ring momentum spectrum with strong interference pattern around at $q_{y}=-1.8m$ in negative $q_{y}$ at the same time, it also interprets why the interference effects at the positive $q_{y}$ plane around $q_{y}=0.9m$ of the inner ring. Because particles created at two different times simply end up at the same final momentum that leads to the enhancing of the total yields, which is the reason why maximum peak appears around at $q_{y}=-1.8m$, see Fig.~\ref{fig:5}($b=20$)

We can understand the interference pattern by considering pair creation as a quantum mechanical process \cite{Blinne:2013via}. Because in the momentum distribution, the right-moving particles along the positive $q_{x}$ direction and the left-moving particles along the negative $q_{x}$ direction carry different phase with the circular distribution. For this circular polarization case, the number of rotation cycles increase with high harmonics $b\omega$, and the ends of the momentum spectrum distribution meet again in negative $q_{y}$ momentum plane and display a ring-shape \cite{Blinne:2013via}. Consequently, the interference term appears in the sum of corresponding quantum mechanical wave functions with different phases. Therefor, we observe interference effects at the negative $q_{y}$ plane.

On the other hand we can also understand the interference structure which change and increase the attribution in the momentum distribution from the semiclassical turning points, shown in Fig. \ref{CP}, for $b\omega=1m$ in the circular polarized $\delta=1$. As we can see, for example when $b\omega=1m$, one of the important difference from the case of circular polarization for the single strong field shown in Fig. \ref{S}(d) is that now the dominant contribution comes from appearance of more turning points. In fact, as $b\omega$ increases, the number of cycles within the pulse duration also increases, thus more turning point pairs come close to the real $t$ axis. Therefore, the interference effects become more and more significant. From the Fig. \ref{CP} one can infer that for $b\omega=1m$ more pairs of turning points appears for $(q_{x}=0,q_{y}=-1.90)$ as compared to $(q_{x}=0,q_{y}=2.25)$. Thus this also could explains why negative momentum in the $q_{y}$ occurs stronger interference effects with smaller width, but higher peaks, see Fig.~\ref{fig:5}($b=20$). In the other words, the calculated results indicate that while the peak value of momentum spectrum in the most negative $q_y$ is higher than that of the most positive $q_y$, the total amount of particles around of the peak at negative case is still lower than that around of the peak at the opposite positive case of $q_y$ plane.

\section{Number density}\label{result2}

\begin{figure}[h]
\begin{center}
\includegraphics[width=10cm]{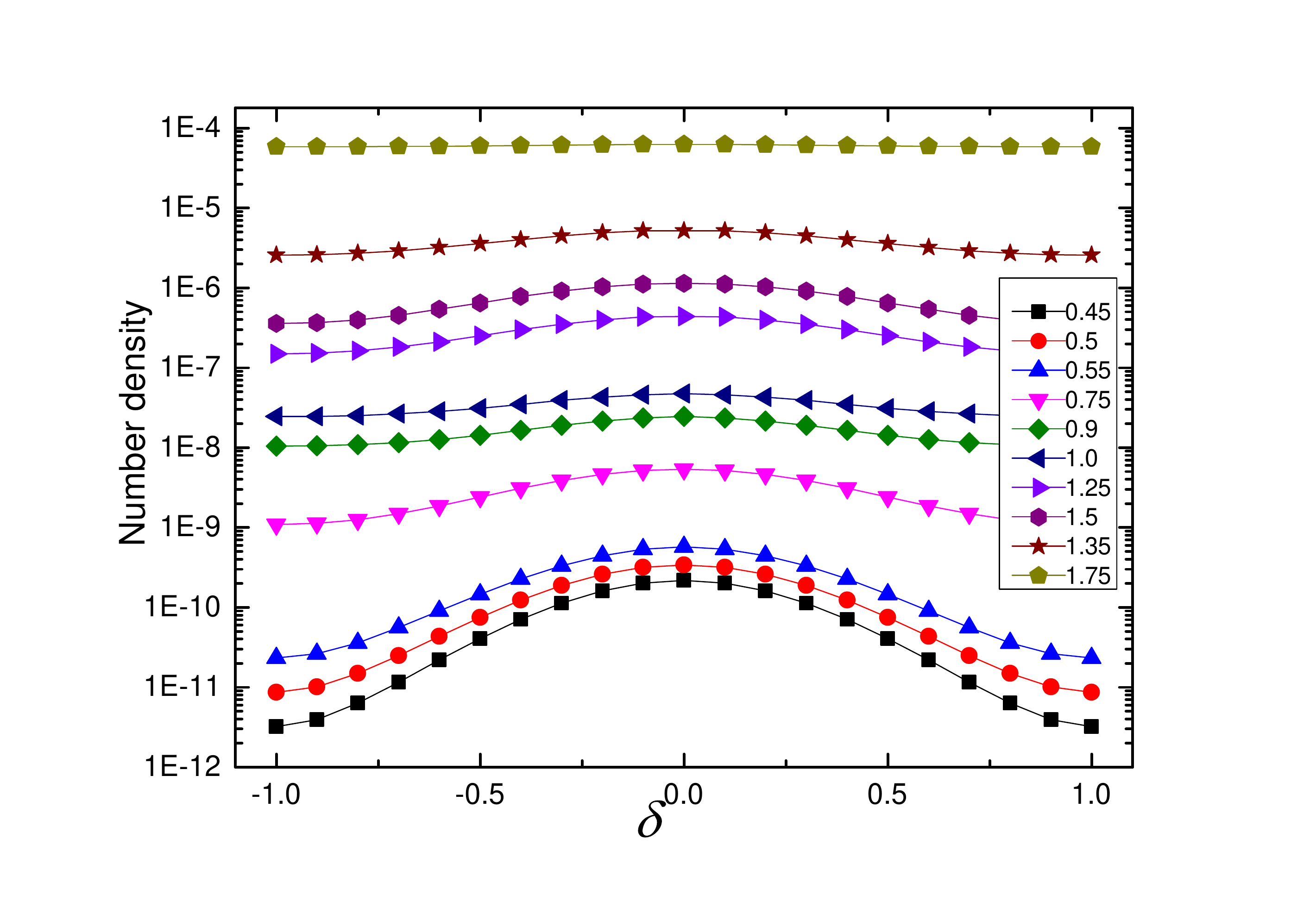}\\
~~\includegraphics[width=10cm]{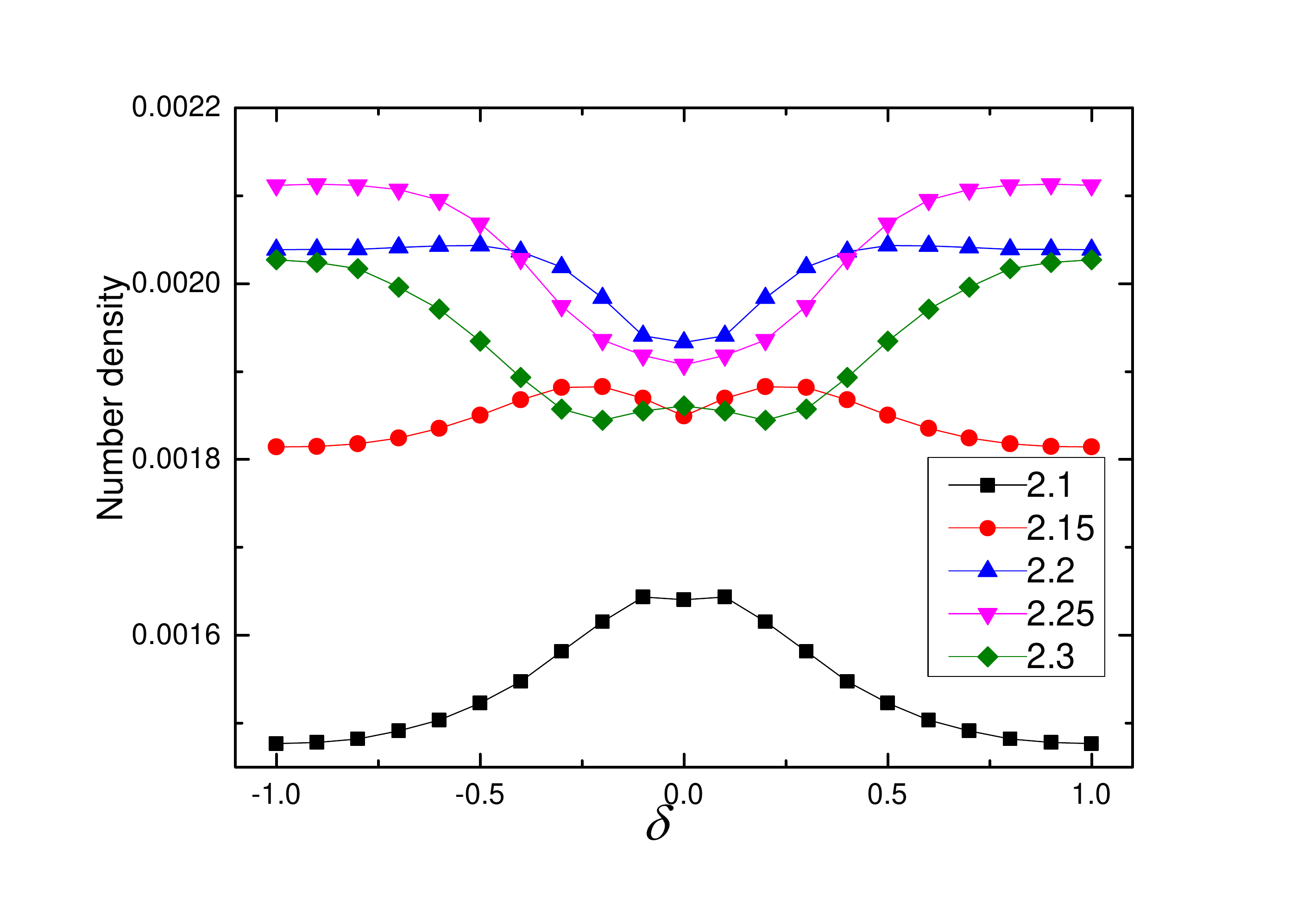}
\end{center}
\vspace{-15mm}
\caption{The number density (in units of $\lambda_{c}^{-3}=m^3$) of created particles as a function of the field polarization $\delta$ for the different high harmonics $b\omega\geq0.45m$. The other field parameters are $E_{1s}=0.1\sqrt{2}E_{cr}$, $E_{2w}=0.01\sqrt{2}E_{cr}$, $\omega=0.05m$, and $\tau=100/m$.}
\label{fig:6}
\end{figure}

The number density is calculated by scanning over the field polarization for different $b\omega$, shown in Fig. \ref{fig:6}. It is seen that the number density enhances as the high harmonics $b\omega$ increases for each polarization, and can reach its maximum at $\delta=0$ until to $b\omega\sim 1.75m$, cf the upper panel of figure. In particular for the small $b\omega$, the number density decreases more sharply compared to the large $b\omega$ as  $\delta$ approaches to $1$.

One possible physical reason is that for small high harmonics $b\omega$, as pair production rate is smaller for
weak fields, the created pairs number density is dominated by the electric-field component with a large peak field strength.
In addition, the polarized electric field given in Eq. \eqref{eq1} with a small value of $\delta\sim0$ has a larger peak strength for the field component compared to the one with a large value of $\delta\sim1$, and therefore results in a high number density.

\begin{figure}[h]\suppressfloats
\includegraphics[width=10cm]{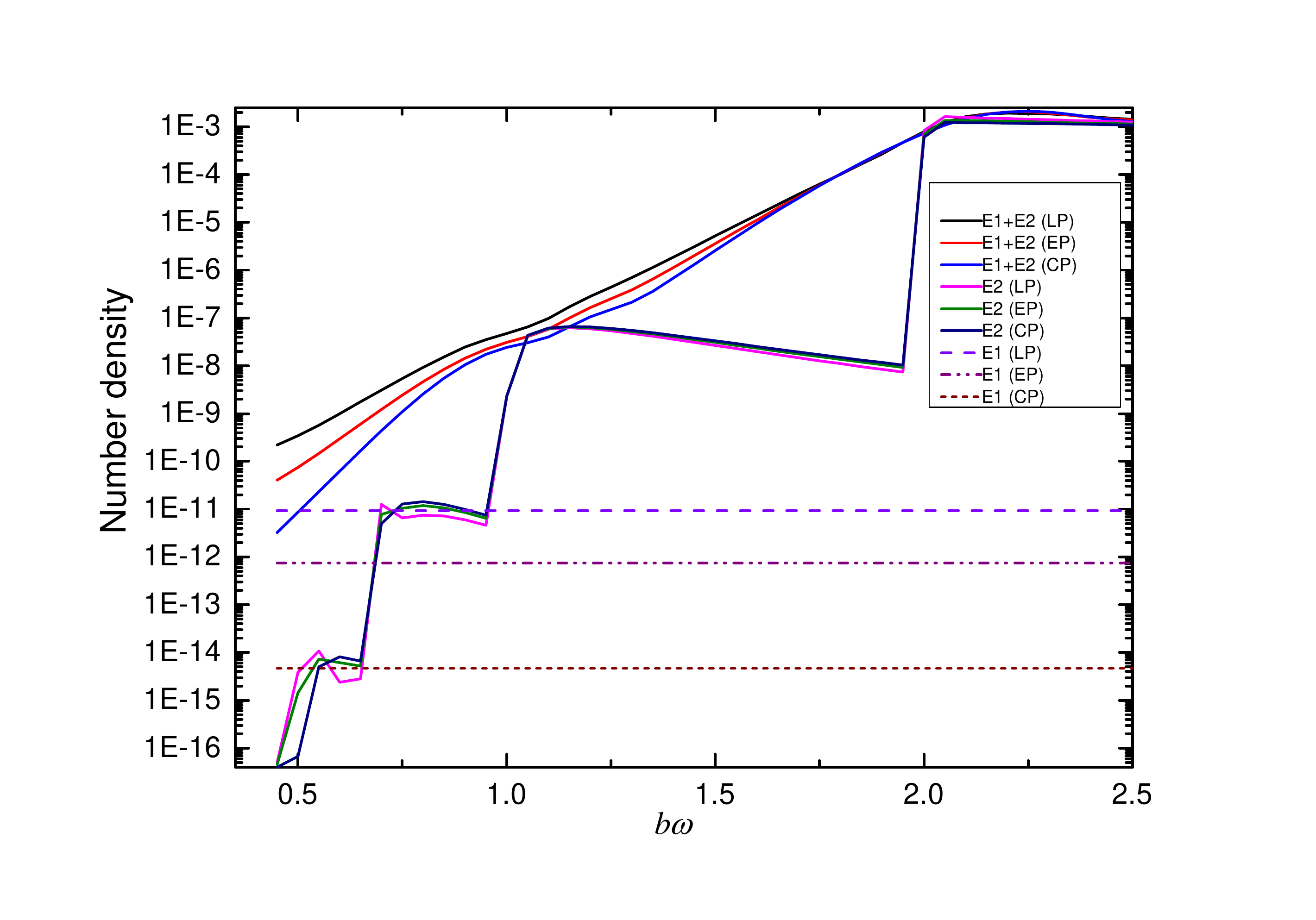}
\caption{The number density (in units of $\lambda_{c}^{-3}=m^3$) of created particles as a function of the high harmonics $b\omega$ for the different polarization $\delta=0$(LP), $\delta=0.5$(EP), and $\delta=1$(CP) for the combined field $E(t)$, strong field $E_{1s}(t)$, and weak field $E_{2w}(t)$, respectively. The other field parameters are $E_{1s}=0.1\sqrt{2}E_{cr}$, $E_{2w}=0.01\sqrt{2}E_{cr}$, $\omega=0.05m$, and $\tau=100/m$.}
\label{fig:7}
\end{figure}

\begin{figure}[h]\suppressfloats
\includegraphics[width=10cm]{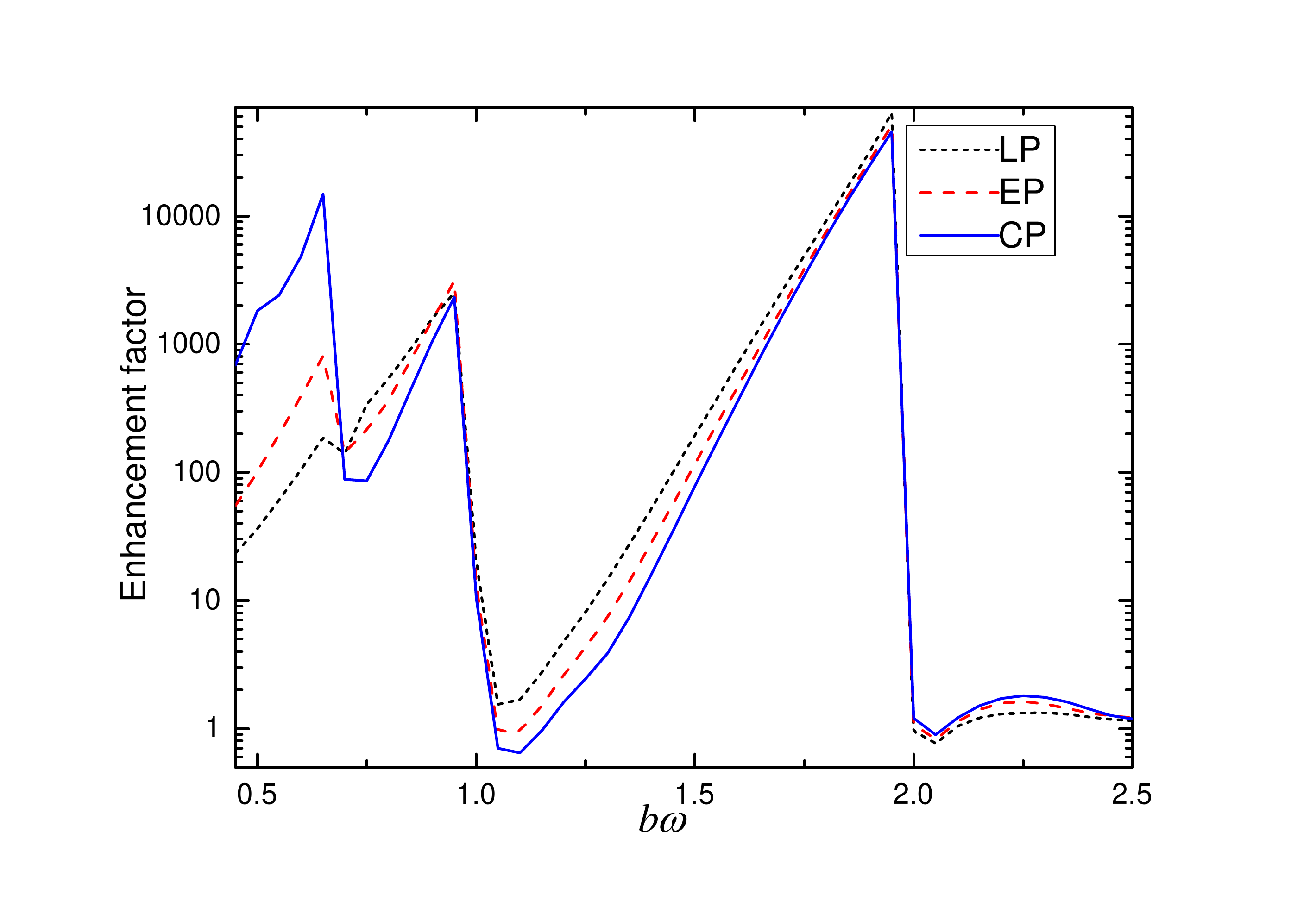}
\caption{The approximate enhancement factor of the number density for each different polarization respectively $\delta=0$(LP), $\delta=0.5$(EP), and $\delta=1$(CP), which is defined by $N_{(1_{s}+2_{w})}/[N_{1_{s}}+N_{2_{w}}]$ as a function of the high harmonics $b\omega$. The other field parameters are $E_{1s}=0.1\sqrt{2}E_{cr}$, $E_{2w}=0.01\sqrt{2}E_{cr}$, $\omega=0.05m$, and $\tau=100/m$.}
\label{fig:9}
\end{figure}

From the lower panel of Fig. \ref{fig:6}, we can find that, for large $2.1m\leq b\omega \leq2.3m$, the relative increment of number density with $b\omega$ is very small and it reaches the maximum at $\delta=0.1$, $0.2$, $0.5$ and $0.9$ instead of either of $\delta=0$ or $1$. For $b\omega=2.3m$, however, its maximum at $\delta=1$ is got again. As a result, for large $b\omega$, we can conclude that the nonlinear relation between the number density and the field polarization.

In order to better understand the relation between the number density and $b\omega$, we depict the number density by scanning over the field high harmonics $b\omega$ for different polarizations in Fig. \ref{fig:7}. For the single strong field from Fig. \ref{fig:7}, we can see that with the increase of the field polarization the number density is monotonically decreasing. In the case of single weak field, for each different polarization, the curves have large leaps at $b\omega\approx2m/n$ indicating the appearance of additional $n$-photon channels.
Due to the effective mass effects of the electron or positron in the strong field, cf. Ref.~\cite{Kohlfurst:2013ura}, it is no surprising that the number density is actually peaked slightly above this expectation $b\omega\approx2m/n$ and this deviation becomes more significant for higher $b\omega$.

From Fig. \ref{fig:7} one can easily infer that for dynamically assisted two-color filed $\mathbf{E}(t)$ a combination of $\mathbf{E}_{1s}(t)$ and $\mathbf{E}_{2w}(t)$ can significantly enhance pair production relative to the separate application of each field for each different polarization. As $b\omega$ increases, the created pairs number density increases for each polarization. When $0.45m\leq b\omega\leq1.75m$, the number density is distinguished for each polarization, and as the polarization parameter $\mid\delta\mid$ increases, the created pairs number density is decreased, therefore, in this region for the linear polarization the produced particle number density is higher than other polarized ones. While in the region $b\omega>1.75m$, we can see that for different polarized fields, the $e^{+}e^{-}$ pair number density is indistinguishable as $b\omega$ further increases. Thus in this region the polarization effect on number density is gradually weaken as $b\omega$. We also see that the high harmonics $b\omega$ for maximum pair production has been shifted to larger at about $b\omega=2.25m$ ($b=45$) compared to the single weak field $\mathbf{E}_{2w}(t)$ (multiphoton) case. Similar effect is also observed in \cite{Jiang} by using the computational quantum field theory approach. Consequently, we can get the optimum number density as in the circular polarization ($\delta=1$) when $b\omega=2.25m$, after that it decreases nonlinearly with $b\omega$.

Finally, in Fig. \ref{fig:9}, we display an approximate enhancement factor $N_{(1_{s}+2_{w})}/[N_{1_{s}}+N_{2_{w}}]$ of the number density for the dynamically assisted two-color filed as function of high harmonics $b\omega$, where $N_{1_{s}}$ and $N_{2_{w}}$ denotes the value of the number density in the case of single strong and single weak field for each different polarizations, respectively, and $N_{(1_{s}+2_{w})}$ is for two-color filed. One can observe that for small $b\omega\leq0.65m$, the enhancement factor is quite distinguished for each polarization with the small enhancement and the enhancement factor increases with the field polarization. While in the region $0.75m\leq b\leq1.95m$, very clearly visible is the suppression in the enhancement factor when one goes from linear to circular polarization and for large values region $1.1m\leq b\leq1.95m$ the enhancement can reach serval orders of magnitude for each polarization. When $b\omega>2m$, the effects of high harmonics parameters $b\omega$ on the approximate enhancement factor is almost negligible for each different polarization.

\section{Summery and Conclusion}

In this study, pair production is studied in dynamically assisted two-color electric fields for different polarization scenarios using the real-time DHW formalism. We have examined features of dynamically assisted pair production in four different situations of linear, near-linear, middle elliptical and circular polarized fields on the momentum spectra and number density of created particles.

For single strong field pulse, the interference effect in the momentum distribution vanishes with the increase of field polarization. While for the dynamically assisted two-color pulse, a strong significant enhancement is seen in the spectrum for the particles cereaed from vacuum when appropriate high harmonics is applied, and interference effects appear with multiple peaks in elliptic polarization as well as in circular polarization. That could be understood semiclassically as the appearance of new turning point pairs due to the additional superimposed weak field. And with the enhancement of high harmonics of two-color fields, interference effects become stronger and the number of multiple peaks also increase for each polarization.

For the number density, it is also found that it exhibits distinctive polarization dependence for small and large high harmonics of two-color fields. When high harmonics is small, the number density decreases with polarization from linear to circular; while for a larger high harmonics, the number density is  sensitive to the high harmonics degree and its polarization dependence exhibits a strong nonlinear characteristic. So the maximal number densities are achieved for either linear, circular or elliptical polarization when different larger high harmonics are applied. It is also noted that for the single strong field, number density decreases with polarization varying from linear to circular. By comparing the number density in each individual field with the combined two-color fields, it is demonstrated that the production rate can be enhanced significantly for each polarization due to the combination of two fields for certain high harmonics parameter region. Finally the polarization effect on number density is gradually weaken as the high harmonics increases while a weak nonlinearity relation appears again if the high harmonics of two-color fields exceeds a single-photon threshold.

We hope that our studies reveal some useful information about pair creation processes in different elliptically polarization scenarios with dynamically assisted two-color field. In this study, we only consider fixed pulse length scale, the results by considering the different pulses with different duration scales would be reported elsewhere.

\begin{acknowledgments}
We thank C. Kohlf\"urst for his critical reading of the manuscript and fruitful discussions. We also thank A. Blinne and M. Ababekri for helpful comments. This work was supported by the National Natural Science Foundation of China (NSFC) under Grant No.\ 11875007 and 11475026.
The computation was carried out at the HSCC of the Beijing Normal University.

\end{acknowledgments}

\end{document}